\documentclass[aps,prc,twocolumn,superscriptaddress,showpacs,showkeys]{revtex4}
\usepackage{hyperref}

\usepackage{graphicx}
\usepackage{dcolumn}
\usepackage{bm}

\usepackage{float}

\usepackage[usenames]{color}

\begin{document}


\title{Competition of coalescence and "fireball" processes in
 nonequilibrium emission of light charged particles
from p+Au collisions }



\author{A.Budzanowski}
\affiliation{H. Niewodnicza{\'n}ski Institute of Nuclear Physics
PAN, Radzikowskiego 152, 31342 Krak{\'o}w, Poland}
\author{M.Fidelus}
\affiliation{M. Smoluchowski Institute of Physics, Jagellonian
University, Reymonta 4, 30059 Krak{\'o}w, Poland}
\author{D.Filges}
\affiliation{Institut f{\"u}r Kernphysik, Forschungszentrum
J{\"u}lich, D-52425 J{\"u}lich, Germany}
\author{F.Goldenbaum}
\affiliation{Institut f{\"u}r Kernphysik, Forschungszentrum
J{\"u}lich, D-52425 J{\"u}lich, Germany}
\author{H.Hodde}
\affiliation{Institut f{\"ur} Strahlen- und Kernphysik, Bonn
University,  D-53121 Bonn, Germany}
\author{L.Jarczyk}
\affiliation{M. Smoluchowski Institute of Physics, Jagellonian
University, Reymonta 4, 30059 Krak{\'o}w, Poland}
\author{B.Kamys}  \email[Electronic address: ]{ufkamys@cyf-kr.edu.pl}
\affiliation{M. Smoluchowski Institute of Physics, Jagellonian
University, Reymonta 4, 30059 Krak{\'o}w, Poland}
\author{M.Kistryn}
\affiliation{H. Niewodnicza{\'n}ski Institute of Nuclear Physics
PAN, Radzikowskiego 152, 31342 Krak{\'o}w, Poland}
\author{St.Kistryn}
\affiliation{M. Smoluchowski Institute of Physics, Jagellonian
University, Reymonta 4, 30059 Krak{\'o}w, Poland}
\author{St.Kliczewski}
\affiliation{H. Niewodnicza{\'n}ski Institute of Nuclear Physics
PAN, Radzikowskiego 152, 31342 Krak{\'o}w, Poland}
\author{A.Kowalczyk}
\affiliation{M. Smoluchowski Institute of Physics, Jagellonian
University, Reymonta 4, 30059 Krak{\'o}w, Poland}
\author{E.Kozik}
\affiliation{H. Niewodnicza{\'n}ski Institute of Nuclear Physics
PAN, Radzikowskiego 152, 31342 Krak{\'o}w, Poland}
\author{P.Kulessa}
\affiliation{H. Niewodnicza{\'n}ski Institute of Nuclear Physics
PAN, Radzikowskiego 152, 31342 Krak{\'o}w, Poland}
\affiliation{Institut f{\"u}r Kernphysik, Forschungszentrum
J{\"u}lich, D-52425 J{\"u}lich, Germany}
\author{H.Machner}
\affiliation{Institut f{\"u}r Kernphysik, Forschungszentrum
J{\"u}lich, D-52425 J{\"u}lich, Germany}
\author{A.Magiera}
\affiliation{M. Smoluchowski Institute of Physics, Jagellonian
University, Reymonta 4, 30059 Krak{\'o}w, Poland}
\author{B.Piskor-Ignatowicz}
\affiliation{M. Smoluchowski Institute of Physics, Jagellonian
University, Reymonta 4, 30059 Krak{\'o}w, Poland}
\affiliation{Institut f{\"u}r Kernphysik, Forschungszentrum
J{\"u}lich, D-52425 J{\"u}lich, Germany}
\author{K.Pysz}
\affiliation{H. Niewodnicza{\'n}ski Institute of Nuclear Physics
PAN, Radzikowskiego 152, 31342 Krak{\'o}w, Poland}
\affiliation{Institut f{\"u}r Kernphysik, Forschungszentrum
J{\"u}lich, D-52425 J{\"u}lich, Germany}
\author{Z.Rudy}
\affiliation{M. Smoluchowski Institute of Physics, Jagellonian
University, Reymonta 4, 30059 Krak{\'o}w, Poland}
\author{R.Siudak}
\affiliation{H. Niewodnicza{\'n}ski Institute of Nuclear Physics
PAN, Radzikowskiego 152, 31342 Krak{\'o}w, Poland}
\affiliation{Institut f{\"u}r Kernphysik, Forschungszentrum
J{\"u}lich, D-52425 J{\"u}lich, Germany}
\author{M.Wojciechowski}
\affiliation{M. Smoluchowski Institute of Physics, Jagellonian
University, Reymonta 4, 30059 Krak{\'o}w, Poland}

\collaboration{PISA - \textbf{P}roton \textbf{I}nduced
\textbf{S}p\textbf{A}llation collaboration}

\date{\today}

\begin{abstract}
The energy and angular dependence of double differential cross
sections $d^{2}\sigma/d\Omega dE$ was measured for
$p,d,t,^{3,4,6}$He, $^{6,7,8,9}$Li, $^{7,9,10}$Be, and
$^{10,11,12}$B  produced in collisions of 1.2 and 1.9 GeV protons
with  Au target.
The beam energy dependence of these data supplemented by the cross
sections from previous experiment at 2.5 GeV is very smooth. The
shape of the spectra and angular distributions almost does not
change in the beam energy range from 1.2 to 2.5 GeV, however, the
absolute value of the cross sections increases for all ejectiles.
The phenomenological model of two emitting, moving sources, with
parameters smoothly varying with energy, reproduces very well
spectra and angular distributions of intermediate mass fragments.
The double differential cross sections for light charged
particles were analyzed in the frame of the microscopic model of
intranuclear cascade with coalescence of nucleons and statistical
model for evaporation of particles from excited residual nuclei.
However, energy and angular dependencies of data agree
satisfactorily neither with predictions of microscopic intranuclear
cascade calculations for protons, nor with coalescence calculations
for other light charged particles.
Phenomenological inclusion of another reaction mechanism - emission
of light charged particles from a "fireball", i.e., fast and hot
moving source - combined with the microscopic model calculations of
intranuclear cascade, coalescence and evaporation of particles leads
to very good description of the data.
It was found that the nonequilibrium processes are very important
for production of light charged particles. They exhaust 40 - 80\% of
the total cross sections - depending on the emitted particles.
Coalescence and "fireball" emission give comparable contributions to
the cross sections with exception of $^{3}$He data where coalescence
clearly dominates.
The ratio of sum of all nonequilibrium processes to those proceeding
through stage of statistical equilibrium does almost not change in
the beam energy range from 1.2 GeV  to 2.5 GeV for all light charged
particles.
\end{abstract}



\maketitle


\section{\label{sec:introduction} Introduction}

In the recent publication \cite{BUB07A} we have shown that the
inclusive spectra of double differential cross sections
$d^{2}\sigma/d\Omega dE$ for light charged particles (LCP's) and
intermediate mass fragments (IMF's) produced in proton - Au
collisions at proton beam energy 2.5 GeV are compatible with the
mechanism similar to cold breakup model proposed by Aichelin et al.
\cite{AIC84A}. According to this model the proton impinging on to
the target drills a cylindrical hole in the nucleus what results in
presence of three sources emitting LCP's, namely a small, fast, and
hot "fireball" consisted of several nucleons \cite{WES76A}, and two
heavier, excited prefragments. They differ significantly in size
because distribution of impact parameters favors non-central
collisions which lead to asymmetric mass values of the products.
Therefore, the heavier prefragment is almost indistinguishable from
the target residuum created in microscopic models as result of the
intranuclear cascade, and the lighter prefragment has typically a
mass of about 20-30 nucleons. IMF's, i.e., the particles heavier
than the $\alpha$ - particle but lighter than fission fragments,
cannot be emitted from the "fireball" because it consists only
several nucleons, however, contributions from both heavier
prefragments have been well visible in their spectra \cite{BUB07A}.

This simple picture of the reaction mechanism is very appealing
because it gives a possibility to understand the presence of large
nonequilibrium contribution to cross sections observed
experimentally, which cannot be quantitatively reproduced by any of
the existing microscopic models based on the assumption of two
stages of the reaction, i.e., the fast stage consisting in
intranuclear cascade of nucleon-nucleon collisions - described by
INC, BUU or QMD models, and the slow stage of reaction in which
heavy target residuum reaches statistical equilibrium and evaporates
particles - described by statistical models.

It should be pointed out, that phenomenological analysis published
in our previous work is not able to unambiguously distinguish
processes proceeding through phase of statistical equilibrium of
heavy target residuum from reactions in which a nonequilibrium
mechanism, i.e., the fast breakup of the target, produces heavy,
excited prefragment moving slowly and therefore being almost
indistinguishable from the target residuum.

To get more insight into the reaction mechanism it is necessary to
investigate energy dependence of the reaction processes as well as
to study interaction of protons with various targets.  A goal of the
present work was to examine beam energy dependence of the emission
of LCP's and IMF's from the collisions of protons with Au target in
a broad proton energy range - from 1.2 GeV to 2.5 GeV. For this
purpose new experimental data were measured and analyzed, confronted
whenever possible with a microscopic description of the data instead
of pure phenomenological treatment as in Ref. \cite{BUB07A}.

Experimental data are discussed in the next section, the theoretical
analysis is described in the third section, discussion of obtained
results is presented in the fourth section, and summary of results
is given in the last section.


\section{\label{sec:experiment}Experimental data}

The experiment was performed with the selfsupporting Au target of
the thickness of about of 300 $\mu$g/cm$^{2}$, irradiated by
internal proton beam of COSY (COoler SYnchrotron) of the J\"ulich
Research Center. The experimental setup and procedure of data taking
were in details described in Refs. \cite{BUB07A} and \cite{BAR04A}.
Thus, here we only point out that the operation of the beam was
performed in so called supercycle mode, i.e. alternating for each
requested beam energy several cycles, consisting of protons
injection to COSY ring, their acceleration with the beam circulating
below the target, and irradiation of the target.
Due to
this all experimental conditions; setup, electronics, the target
thickness and its position were exactly the same for all three
studied proton energies - 1.2, 1.9 and 2.5 GeV. In this way the
energy dependence was not biased by systematic effects caused by
possible modifications of the experimental conditions for
experiments with different beam energy.

\begin{figure}
\begin{center}
\includegraphics[angle=0,width=0.5\textwidth]{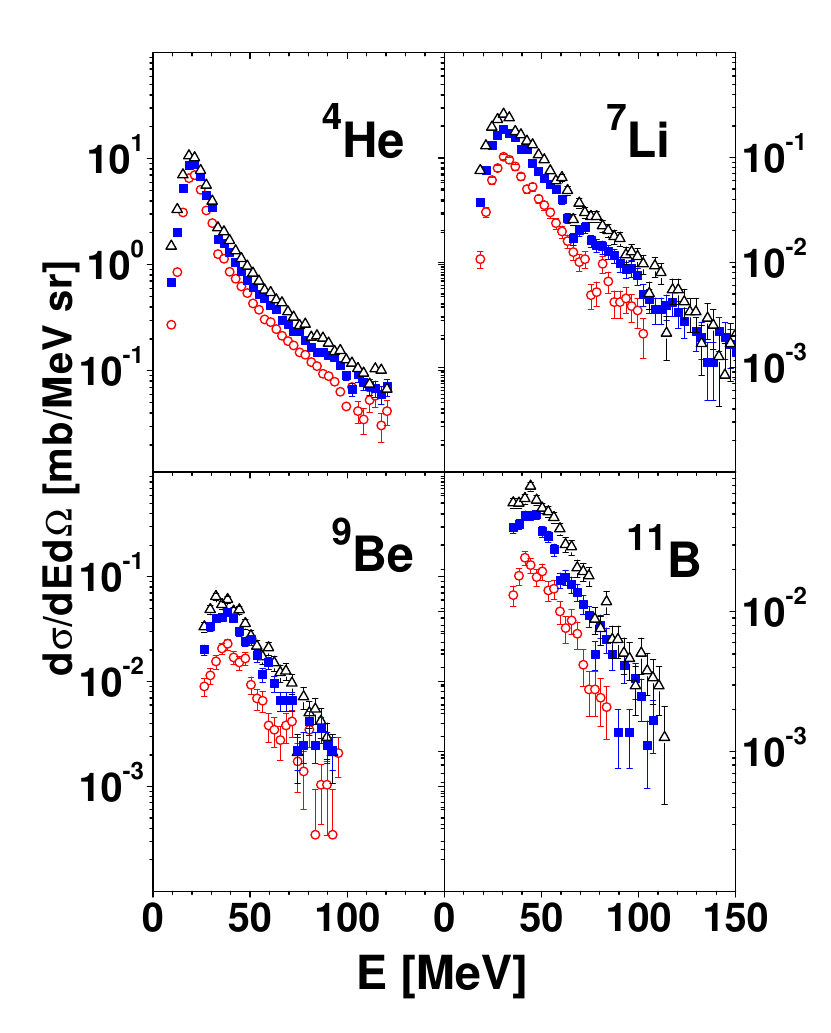}
\caption{\label{fig:helibeb} (Color online) Typical spectra of
$^{4}$He,  $^{7}$Li, $^{9}$Be, and $^{11}$B ejectiles  (upper left,
upper right, lower left, and lower right parts of the figure,
respectively) measured at 35$^{\circ}$ for three energies of the
proton beam; 1.2, 1.9, and 2.5 GeV, impinging on to the Au target.
Open circles represent the lowest energy, full squares -  the
intermediate energy, whereas open triangles show the data for the
highest energy. The cross sections at 2.5 GeV proton beam energy
were published in Ref. \cite{BUB07A} and the data at 1.2 and 1.9 GeV
were obtained in the present experiment.}
\end{center}
\end{figure}

Double differential cross sections $d^{2}\sigma/d\Omega dE$ were
measured as a function of scattering angle and energy of ejectiles,
which were mass and charge identified for isotopes of H, He, Li, Be,
and B.  Heavier ejectiles i.e. C, N, O, F, Ne, Na, Mg, and Al were
only charge identified.
Typical spectra of isotopically identified ejectiles obtained in the
present experiment are shown in Fig. \ref{fig:helibeb}. As can be
seen in this figure  the shape of spectra does not vary
significantly with beam energy. The main effect, present for all
products is monotonic increase of the absolute value of the cross
sections with beam energy. Furthermore, all the spectra contain two
components; low energy component of the Gaussian shape - attributed
to evaporation from an equilibrated, excited nucleus, and high
energy exponential component - interpreted as nonequilibrium
mechanism contribution. The data for LCP's, represented in Fig.
\ref{fig:helibeb} by $\alpha$-particles, have similar character and
energy dependence as those for IMF's, however, the nonequilibrium
component is more pronounced.


\section{\label{sec:analysis} Theoretical analysis}

The equilibrium emission of LCP's and IMF's may be portrayed by
statistical model of particle evaporation from excited heavy target
residuum created in the fast stage of the reaction.
 This is, however, not the case for
nonequilibrium emission of composite particles, which cannot be
satisfactorily  described by models used for reproduction of the
first stage of the reaction, i.e., by intranuclear cascade,
Boltzmann-Uehling-Uhlenbeck or Quantum Molecular Dynamics models.
All mentioned models of the reaction neglect to large extent
possible multinucleon correlations, which can be crucial for
nonequilibrium processes. Whereas it is possible to take effectively
these correlations into account for LCP's -  by introducing
coalescence of emitted nucleons into clusters - such a procedure is
not sufficient for description of IMF's nonequilibrium emission.
From this reason different theoretical analysis has been performed
for LCP's and for IMF's.


%
\begin{table*}
  \caption{\label{table:parameters}Parameters of two moving sources for isotopically identified
  IMF's: k, $\beta$, T, and $\sigma$ correspond to reduced height
  of the Coulomb barrier for emission of fragments
  (see the text for the explanation), source velocity, its apparent temperature,
 and total (integrated over angle and energy of detected particles)
 production cross section, respectively.  The left part of the Table
 (parameters with indices "1")
 corresponds to the slow moving source, and the right part of the
 Table \ref{table:parameters} contains values of parameters for the fast moving source.
  The upper row for each ejectile corresponds to beam energy 1.2 GeV, the intermediate row to 1.9 GeV,
  and the lowest one to the energy 2.5 GeV.  Velocities for slow sources are fixed at value 0.003c estimated
  as velocities of heavy target residua from intranuclear cascade calculations.
   }
\begin{tabular}{llrr|lllrc}
  \hline \hline
            & \multicolumn{3}{c|}{Slow source} & \multicolumn{4}{c}{Fast source}                 &  \\
  \cline{2-8}
  Ejectile  & \emph{k}$_1$    & \emph{T}$_1$/MeV & $\sigma_1$/mb & \emph{k}$_2$  &     $\beta_2$   & \emph{T}$_2$/MeV         & $\sigma_2$/mb & $\chi^2$ \\
  \hline
  $^6$He    & 0.97 $\pm$ 0.09 &  9.3 $\pm$ 1.1 &  8.1 $\pm$ 1.1 & 0.47 $\pm$ 0.05 & 0.034 $\pm$ 0.007   & 13.6 $\pm$ 1.2 &  3.9 $\pm$ 1.0 & 2.7 \\
            & 0.95 $\pm$ 0.04 &  9.1 $\pm$ 0.6 & 18.5 $\pm$ 1.2 & 0.36 $\pm$ 0.05 & 0.040 $\pm$ 0.007   & 19.1 $\pm$ 1.3 &  4.9 $\pm$ 1.1 & 2.6 \\
            & 0.97 $\pm$ 0.04 &  9.0 $\pm$ 0.6 & 24.8 $\pm$ 1.4 & 0.35 $\pm$ 0.05 & 0.040 $\pm$ 0.007   & 21.6 $\pm$ 1.4 &  7.5 $\pm$ 1.4 & 2.1 \\
  \hline
  $^6$Li    & 0.89 $\pm$ 0.05 & 12.4 $\pm$ 0.9 & 10.5 $\pm$ 0.8 & 0.43 $\pm$ 0.08 & 0.047 $\pm$ 0.008   & 22.2 $\pm$ 1.3 & 2.85 $\pm$ 1.3 & 2.4\\
            & 0.85 $\pm$ 0.04 & 12.1 $\pm$ 0.7 & 19.5 $\pm$ 1.2 & 0.43 $\pm$ 0.05 & 0.040 $\pm$ 0.004   & 23.6 $\pm$ 0.7 &  7.7 $\pm$ 1.1 & 2.4\\
            & 0.86 $\pm$ 0.04 & 11.1 $\pm$ 0.8 & 25.3 $\pm$ 1.7 & 0.44 $\pm$ 0.04 & 0.034 $\pm$ 0.003   & 23.7 $\pm$ 0.6 & 14.5 $\pm$ 1.7 & 2.0\\
  \hline
  $^7$Li    & 0.89            & 12.3           & 18.0           & 0.47            & 0.039               & 16.4           & 4.8            & 4.7\\
            & 0.88 $\pm$ 0.03 & 11.7 $\pm$ 0.5 & 38.1 $\pm$ 1.8 & 0.37 $\pm$ 0.04 & 0.040 $\pm$ 0.005   & 20.3 $\pm$ 0.7 & 10.3 $\pm$ 1.7 & 4.2\\
            & 0.88 $\pm$ 0.03 & 11.6 $\pm$ 0.6 & 50.8 $\pm$ 2.6 & 0.36 $\pm$ 0.03 & 0.035 $\pm$ 0.003   & 20.9 $\pm$ 0.5 & 20.3 $\pm$ 2.6 & 3.1\\
  \hline
  $^8$Li    & 0.94 $\pm$ 0.11 & 11.1 $\pm$ 1.6 & 3.51 $\pm$ 0.45& 0.48 $\pm$ 0.08 & 0.040 $\pm$ 0.008   & 14.4 $\pm$ 2.0 & 1.15 $\pm$ 0.45 & 1.8\\
            & 0.90 $\pm$ 0.08 & 11.8 $\pm$ 1.3 & 6.65 $\pm$ 0.90& 0.43 $\pm$ 0.05 & 0.032 $\pm$ 0.006   & 17.2 $\pm$ 1.1 & 3.65 $\pm$ 0.93 & 2.5\\
            & 0.90 $\pm$ 0.09 & 11.9 $\pm$ 1.5 &  9.1 $\pm$ 1.4 & 0.45 $\pm$ 0.05 & 0.029 $\pm$ 0.005   & 18.0 $\pm$ 1.0 &  6.4 $\pm$ 1.5  & 2.1\\
  \hline
  $^9$Li    & 1.01 $\pm$ 0.19 & 11.9 $\pm$ 2.9 & 0.92 $\pm$ 0.09 & 0.58 $\pm$ 0.33 & 0.044 $\pm$ 0.008   &  4.1 $\pm$ 1.8 &  0.25 $\pm$ 0.12 & 1.1\\
            & 0.84 $\pm$ 0.09 & 10.4 $\pm$ 3.0 & 1.92 $\pm$ 0.37 & 0.51 $\pm$ 0.08 & 0.034 $\pm$ 0.008   & 11.9 $\pm$ 2.5 &  0.77 $\pm$ 0.33 & 1.5\\
            & 1.00 $\pm$ 0.22 & 10.4 $\pm$ 3.0 &  2.1 $\pm$ 0.5  & 0.39 $\pm$ 0.07 & 0.025 $\pm$ 0.003   & 18.2 $\pm$ 1.6 &   2.1 $\pm$ 0.6  & 1.2\\
  \hline
  $^7$Be    & 0.89            & 13.3           &  1.22          & 0.52            & 0.036               & 25.3           &  0.88          & 1.1\\
            & 0.86 $\pm$ 0.21 & 14.1 $\pm$ 5.3 &  1.7 $\pm$ 1.0 & 0.61 $\pm$ 0.06 & 0.025 $\pm$ 0.007   & 22.8 $\pm$ 1.2 &  2.9 $\pm$ 1.0 & 1.2\\
            & 0.92 $\pm$ 0.27 & 11.2 $\pm$ 4.3 &  2.6 $\pm$ 0.8 & 0.48 $\pm$ 0.05 & 0.038 $\pm$ 0.005   & 24.0 $\pm$ 1.2 &  4.6 $\pm$ 0.9 & 1.4\\
  \hline
  $^9$Be    & 0.86            &  9.7           & 5.2            & 0.50            & 0.030               & 15.2           &  1.24          & 1.7\\
            & 0.88            &  9.8           & 9.5            & 0.59            & 0.022               & 15.0           &  4.41          & 1.4\\
            & 0.86 $\pm$ 0.12 &  9.6 $\pm$ 1.7 & 12.5 $\pm$ 1.9 & 0.53 $\pm$ 0.06 & 0.020 $\pm$ 0.005   & 16.6 $\pm$ 0.8 &  8.1 $\pm$ 2.3 & 1.4\\
  \hline
  $^{10}$Be & 0.86 $\pm$ 0.16 & 12.4 $\pm$ 2.1 & 3.5  $\pm$ 1.3 & 0.62 $\pm$ 0.14 & 0.024 $\pm$ 0.011   & 9.0  $\pm$ 3.7 &  1.9 $\pm$ 1.3 & 1.8\\
            & 0.86            & 12.0           & 7.34           & 0.47            & 0.027               & 13.3           &  13.3          & 1.8\\
            & 0.90 $\pm$ 0.08 & 11.8 $\pm$ 1.2 & 10.0 $\pm$ 1.4 & 0.44 $\pm$ 0.04 & 0.026 $\pm$ 0.004   & 14.5 $\pm$ 0.9 &  6.8 $\pm$ 1.5 & 1.3\\
  \hline
  $^{10}$B  & 0.83            & 11.7           & 1.61           & 0.78            & 0.017               & 15.9           &  0.83          & 2.8\\
            & 0.87            & 10.2           & 4.93           & 0.70            & 0.021               & 17.7           &  1.64          & 1.5\\
            & 0.85 $\pm$ 0.20 & 10.5 $\pm$ 3.4 & 6.6  $\pm$ 1.3 & 0.73 $\pm$ 0.14 & 0.020 $\pm$ 0.010   & 18.2 $\pm$ 2.7 &  2.7 $\pm$ 1.7 & 1.8\\
  \hline
  $^{11}$B  & 0.84 $\pm$ 0.11 & 10.6 $\pm$ 1.4 & 5.9  $\pm$ 0.7 & 0.53 $\pm$ 0.08 & 0.032 $\pm$ 0.008   & 10.6 $\pm$ 2.2 & 1.9  $\pm$ 0.6 & 0.94\\
            & 0.90            & 10.2           & 8.2            & 0.57            & 0.019               & 13.9           & 8.7            & 1.6\\
            & 0.93 $\pm$ 0.18 & 10.5 $\pm$ 2.1 & 12.8 $\pm$ 2.5 & 0.50 $\pm$ 0.05 & 0.022 $\pm$ 0.004   & 14.5 $\pm$ 0.7 & 12.8 $\pm$ 2.8 & 1.7\\
  \hline
  $^{12}$B  & 0.83            &  11.9          & 1.39           & 0.54            & [0.032]             & 12.5           &  0.46          & 1.3\\
            & 0.88            &  7.8           & 2.12           & 0.71            & 0.017               & 13.5           &  2.57          & 1.2\\
            & 0.87            &  8.8           & 1.6            & 0.73            & 0.012               & 13.2           &  5.1           & 1.0\\
  \hline
  \hline
\end{tabular}
\end{table*}
%
The IMF's data have been analyzed in the frame of phenomenological
model of two moving sources as it was done for the data measured at
2.5 GeV beam energy in the previous investigation of these reactions
\cite{BUB07A}. In this way the energy dependence of IMF's production
could be studied in a consistent way.  This
analysis is described in subsection \ref{sec:IMF}. 

The LCP's nonequilibrium emission can be, on the contrary, analyzed
in the frame of the microscopic model, which assumes that the
mechanism of nonequilibrium reactions consists in intranuclear
cascade of nucleon-nucleon collisions \cite{BOU02A} accompanied by
coalescence of the nucleons escaping from the nucleus as it was done
in Refs. \cite{LET02A},\cite{BOU04A}. The authors of these papers
claimed that the main properties of nonequilibrium emission of LCP's
are well reproduced by the proposed microscopic model. 
 Thus, in the
present study the INCL4.3 computer program \cite{BOU04A} has been
used for description of the intranuclear cascade of nucleon-nucleon
collisions with inclusion of coalescence of nucleons, whereas the
GEM2 computer program \cite{FUR00A},\cite{FUR02A} served for
evaluation of evaporation of particles from heavy target residuum
remaining after the intranuclear cascade.  It was also investigated
whether eventual disagreement of the microscopic model calculations
with experimental results leaves still a room for contribution from
another mechanism, namely the "fireball" emission postulated in our
previous paper \cite{BUB07A}. This analysis is
described in subsection \ref{sec:LCP}. 


\subsection{\label{sec:IMF} Intermediate mass fragments}

The main assumptions of the phenomenological model of two moving
sources have been formulated in the paper of Westfall \emph{et al.}
\cite{WES78A}.  They consist in description of double differential
cross sections $d^{2}\sigma/d\Omega dE$ as incoherent sum of
contributions originating from isotropic emission of particles from
two sources moving in direction parallel to the beam direction. Each
of the sources has Maxwellian distribution of the energy available
for the two body decay resulting in emission of the detected
particles.  Velocity of the source - $\beta$, its temperature - $T$,
and contribution to the total production cross section - $\sigma$
are treated as free parameters. The presence of the Coulomb barrier
which hinders emission of low energy particles was originally taken
into account by energy sharp cut off, smoothed in turn by weighting
with uniform or Gaussian probability distribution of the height of
the barrier. In our recent paper \cite{BUB07A} we used another
method, namely we multiplied the Maxwellian energy distribution by
smooth function corresponding to transmission probability through
the barrier. Presence of Coulomb barrier introduces two parameters
which influence mainly low energy part of the spectra:
$k$-parameter, i.e., height of the Coulomb barrier in units of the
height of barrier B of two charged, touching spheres of radius 1.44
A$^{1/3}$;  B~=~Z$_1$~Z$_2$~e$^2$/1.44 (A$_{1}^{1/3}$ +
A$_{2}^{1/3}$), and ratio B/d , where d is a diffuseness of the
transmission function through the barrier:
$P(E)=(1+exp((E-kB)/d)))^{-1}$. Details of this procedure, as well
as interpretation of parameters of the model can be found in
Appendix of Ref. \cite{BUB07A}.

\begin{figure}
\begin{center}
\includegraphics[angle=0,width=0.5\textwidth]{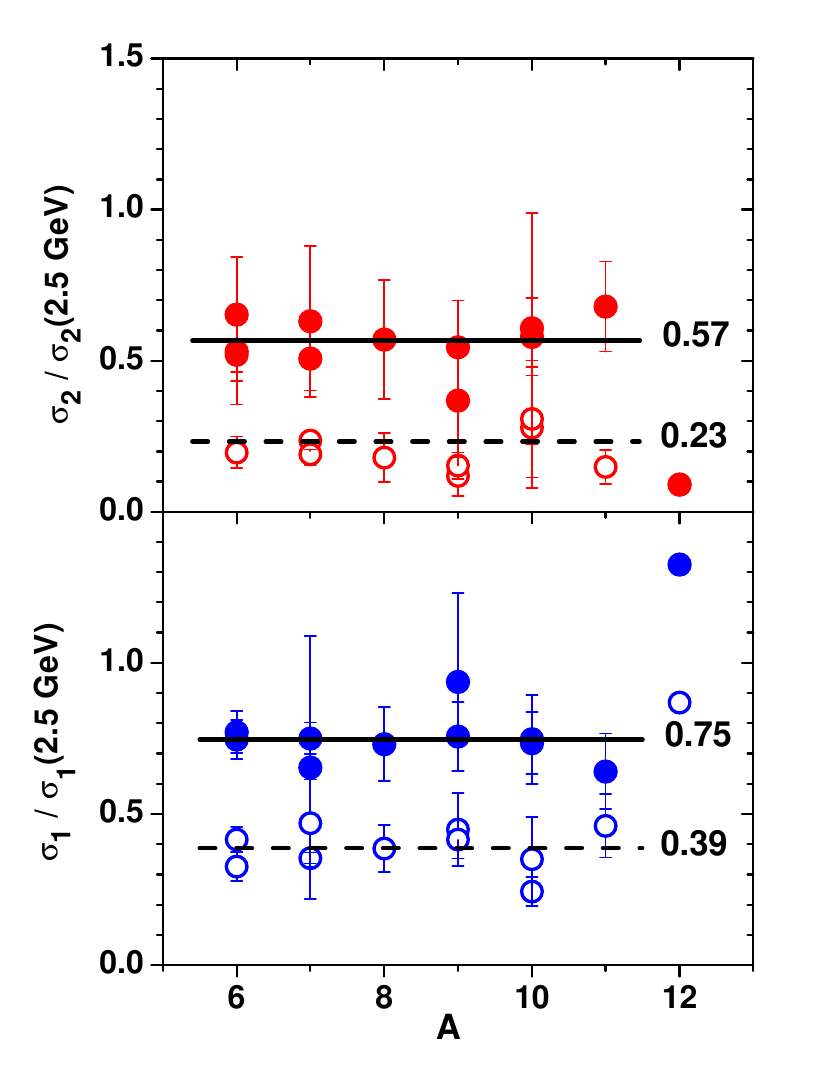}
\caption{\label{fig:r12imfoda} (Color online) Symbols $\sigma_1$ and
$\sigma_2$ correspond to slow and fast emitting source,
respectively. Full dots represent ratio of production cross sections
at beam energy 1.9 GeV to those found at 2.5 GeV as a function of
mass of emitted IMF's. Open circles depict such a ratio for cross
sections measured at 1.2 GeV to those determined at 2.5 GeV.
The lines, present in the figure show average
values of the ratios: 0.23, and 0.57 for the fast source at 1.2 GeV,
and 1.9 GeV, respectively, as well as 0.39, and 0.75 for the slow
source at these energies.}
\end{center}
\end{figure}

The parameters of two moving sources were fitted to experimental
data consisted of energy spectra measured at seven angles:
16$^{\circ}$, 20$^{\circ}$, 35$^{\circ}$, 50$^{\circ}$,
65$^{\circ}$, 80$^{\circ}$, and 100$^{\circ}$. To decrease the
number of parameters it was assumed that velocity of the slow source
emitting IMF's is equal to velocity of the heavy residuum from
intranuclear cascade, i.e., $\beta_{1}$=0.003. Variation of this
velocity influences very slightly values of other parameters, e.g.,
its modification by 30\% causes changes of other parameters smaller
than their errors estimated by fitting computer program.
Furthermore, the $B/d$ ratio was arbitrarily assumed to be equal to
5.5. In evaluation of k-parameter it was assumed that B is defined
as the Coulomb barrier between the emitted particles and the target
nucleus.  This assumption allows for easy comparison of k-parameter
values for different ejectiles and emitting sources.

The computer program searching for the best fit values of the
parameters was able in most cases to provide estimation of errors of
the parameters.  However, sometimes this was not possible,
especially when strong ambiguities of parameters were present.
Therefore, some values of the parameters are quoted without
estimation of errors.  In this case it may happen that the accuracy
of determination of this parameters is poorer then that for the
parameters accompanied by estimates of errors.

Very good description of the spectra of all IMF's has been obtained
as can be judged from $\chi^{2}$ values quoted in the Table
\ref{table:parameters}, which vary usually between 1 and 2.

\begin{figure}
\begin{center}
\includegraphics[angle=0,width=0.5\textwidth]{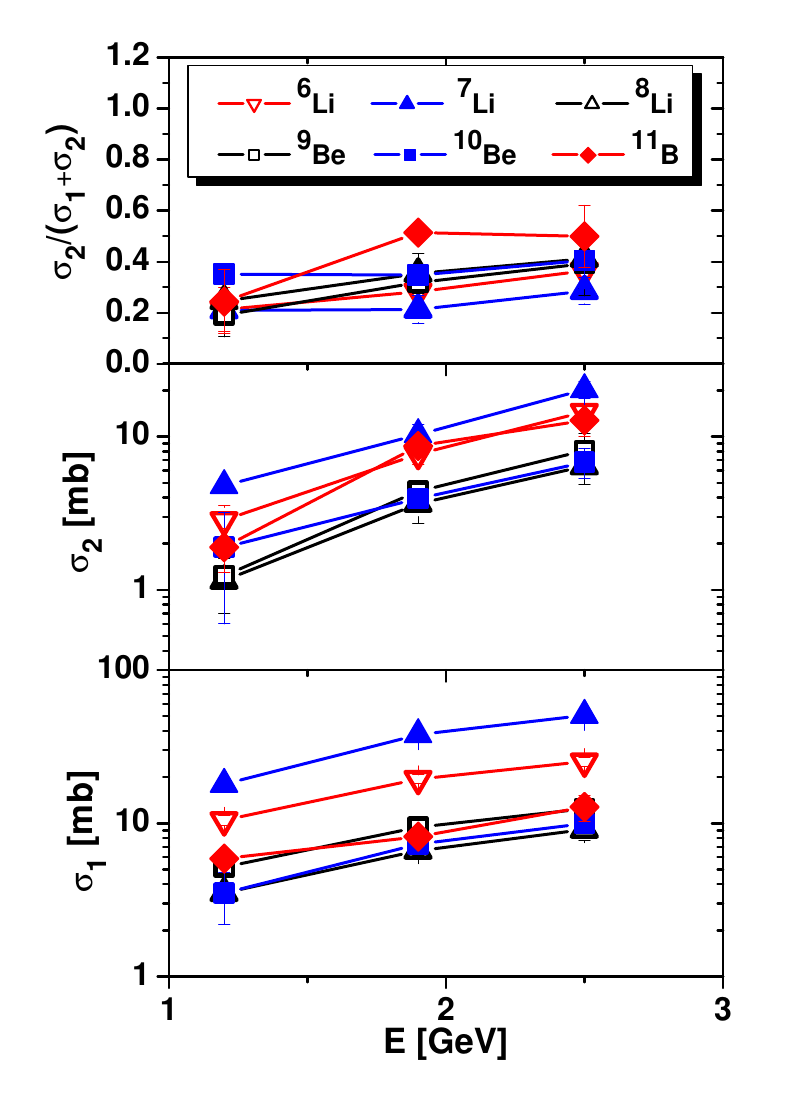}
\caption{\label{fig:imf2to12vse} (Color online) Energy dependence of
the cross section $\sigma_1$, corresponding to emission from the
slow source, is shown in the lower part of the figure, energy
dependence of the cross section $\sigma_2$, related to emission from
the fast source, is depicted  in the middle part of the figure,
whereas energy dependence of the relative contribution of the fast
source is presented in the upper part of the figure.}
\end{center}
\end{figure}

As can be seen from the Table \ref{table:parameters}, values of the
parameters found from the fit to the data obtained at 1.2 and at 1.9
GeV are very close to those which were determined in the analysis of
the data at 2.5 GeV beam energy. This is not true for the total
cross sections which increase monotonically with energy for both
emitting sources.  This increase is illustrated by Fig.
\ref{fig:r12imfoda} where ratios of the total cross sections found
for data at 1.2 GeV, and at 1.9 GeV to cross sections found for data
at 2.5 GeV are shown as open circles and full dots respectively. The
ratios of total cross sections for the fast source are shown in the
upper part of the figure and those for the slow source are depicted
in the lower part of the figure. The following conclusions can be
derived from inspection of Fig. \ref{fig:r12imfoda} :
\begin{description}
  \item[  (i)] The ratios of the cross sections for both sources $\sigma_1(E,A)/\sigma_1(2.5 GeV,A)$ and
  $\sigma_2(E,A)/\sigma_2(2.5 GeV,A)$
   are independent of the
  mass A of ejectiles (with exception of the $^{12}$B cross sections,
  which are, however, not well determined because of poor statistics
  of the data).
  \item[ (ii)] Cross sections for both sources are always larger for E=1.9 GeV  than
  cross sections for E=1.2 GeV (full dots are above open circles)
  and cross sections for E=2.5 GeV are the largest (the ratios are
  always smaller than unity).
  \item[(iii)] The averaged over mass of ejectiles ratios of the cross sections for the slow
  source, i.e. $<~\sigma_1($1.2~GeV$)~/~\sigma_1(2.5$ GeV)$~> =0.39$,
  and $<~\sigma_1($1.9 GeV$)~/~\sigma_1(2.5$ GeV)$~> =0.75$,
  are larger than the corresponding ratios for the fast source, i.e.
  $<~\sigma_2($1.2 GeV$)~/~\sigma_2(2.5$ GeV$)~> =0.23$,
  and $<~\sigma_2($1.9 GeV$)~/~\sigma_2(2.5$ GeV$)> =0.57$.
  This means that the cross sections of the slow
  source increase relatively slower in the beam energy range from 1.2 GeV
  to 2.5 GeV than the cross sections attributed to the fast source,
  thus the contribution from the fast source
  becomes more important for higher beam energy.
  This is confirmed by the fact, that the relative contribution
  $\sigma_2(E,A)/(\sigma_1(E,A)+\sigma_2(E,A))$
  of the fast source  to the total production cross section of IMF's, evaluated
  using the numbers from the Table \ref{table:parameters},
  increases with energy in almost the same way for all IMF's.
  In average, this contribution is equal to 0.27~$\pm$~0.03,
  0.33~$\pm$~0.05, and 0.44 $\pm$ 0.05 for beam energy equal to 1.2,
  1.9, and 2.5 GeV, respectively.

  The above findings are also illustrated by Fig.
  \ref{fig:imf2to12vse} in which energy dependence of cross sections
  $\sigma_1$ and $\sigma_2$ is shown for emission from the slow source
  and fast source, respectively, as well as energy dependence of the
  relative contribution of the fast source $\sigma_2$ to the total cross
  section $\sigma_1 + \sigma_2$. Note using of different scales; linear for
  the upper part of the figure, and logarithmic for the middle and lower
  parts of the figure. It may be observed, that $\sigma_1$ and $\sigma_2$,
  vary rather fast with energy; $\sigma_1$ increases $\sim$ 2-3 times
  in the studied energy range whereas $\sigma_2$ increases even more, i.e., $\sim$ 3-5 times.
  However the relative contribution of nonequilibrium mechanism,
  i.e., $\sigma_2/(\sigma_1+\sigma_2)$  increases much slower,
  as it was mentioned above, because of the same energy trend for both cross
  sections $\sigma_1$ and $\sigma_2$.
\end{description}

\subsection{\label{sec:LCP} Light charged particles}

It is well known that the cross sections for production of LCP's are
at least order of magnitude larger than cross sections for emission
of IMF's. Therefore, knowledge of the mechanism of LCP's production
is crucial for understanding of the full interaction process. The
coalescence mechanism seems to be very promising for explanation of
nonequilibrium production of LCP's \cite{LET02A},\cite{BOU04A}.
However, it is obvious, that such a hypothesis relies on the proper
reproduction of the nucleon spectra by intranuclear cascade
mechanism.
In the case of lack of good description of the proton spectra, the
coalescence mechanism cannot alone be responsible for the observed
nonequilibrium emission of LCP's. To study importance of the
coalescence in the production of LCP's, the experimental proton
spectra for three studied energies were compared with predictions of
the intranuclear cascade model coupled with evaporation of nucleons.
The calculations have been performed by means of INCL4.3 computer
program \cite{BOU04A} in which coalescence of nucleons can be taken
optionally into account, whereas evaporation of protons as well as
complex particles was described by GEM2 computer program
\cite{FUR00A},\cite{FUR02A}.

\begin{figure}
\begin{center}
\includegraphics[angle=0,width=0.5\textwidth]{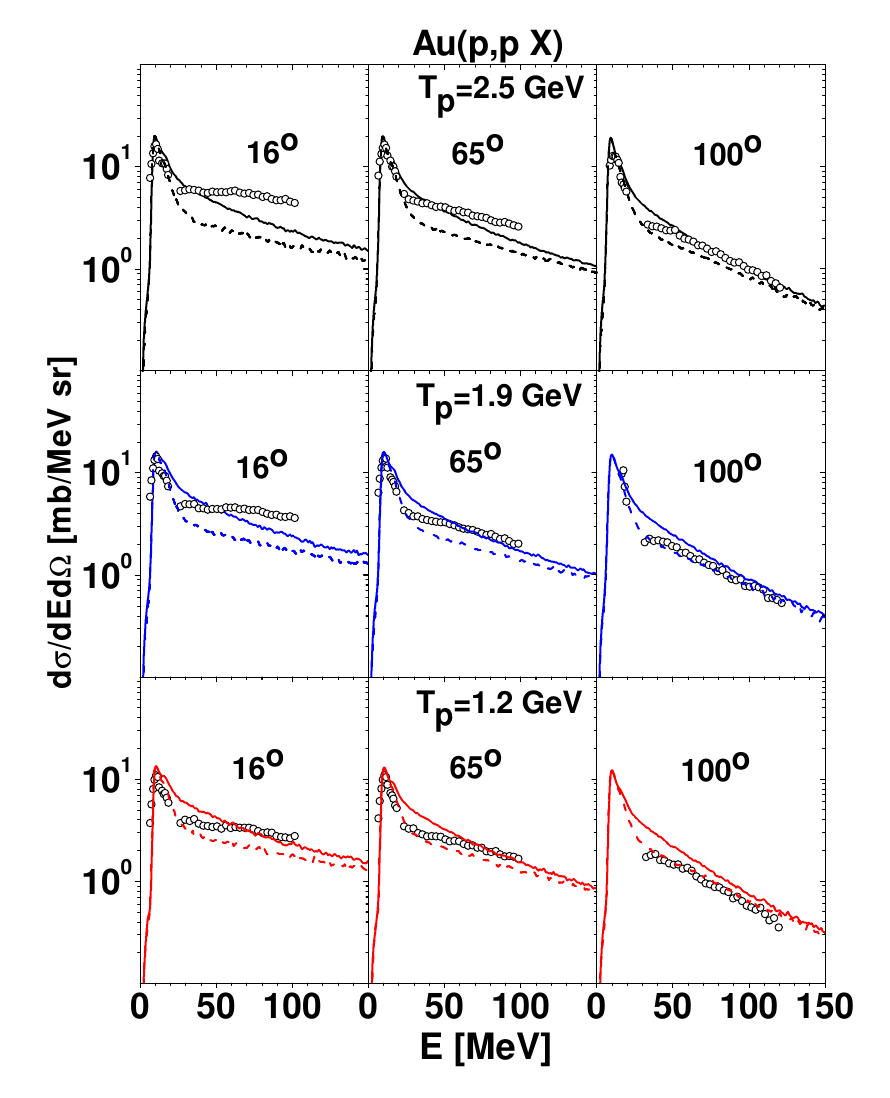}
\caption{\label{fig:au234icg} (Color online) Open circles represent
experimental energy spectra of protons measured at selected angles:
16$^{\circ}$, 65$^{\circ}$, and 100$^{\circ}$ (left, central, and
right column of the figure, respectively)  for three proton beam
energies: 1.2, 1.9 GeV -- present experiment , and 2.5 GeV -- Ref.
\cite{BUB07A} (lower, central, and upper row of the figure,
respectively). The solid lines show results of calculations
performed in the frame of intranuclear cascade  formalism by means
of INCL4.3 program \cite{BOU04A} combined with the evaporation of
protons from excited residual nuclei after fast stage of the
reaction evaluated by means of the GEM2 program of S. Furihata
\cite{FUR00A},\cite{FUR02A}. The dashed lines present calculations
made also with INCL4.3 plus GEM2 programs, however, the coalescence
of nucleons into light complex particles is taken into account
according to prescription proposed in Ref. \cite{BOU04A}.}
\end{center}
\end{figure}

Such calculations, done with inclusion of coalescence and without
this mechanism, are presented in Fig. \ref{fig:au234icg} as dashed
and solid lines, respectively, together with experimental proton
spectra - circles.

As can be seen, the theoretical spectra obtained from calculations
neglecting the coalescence overestimate the experimental spectra at
proton beam energy of 1.2 GeV, but underestimate big part of the
spectra at beam energy of 2.5 GeV in particular for most forward
angles. It seems, that the theoretical proton spectra evaluated
without coalescence have different beam energy dependence then the
experimental data. Inclusion of coalescence significantly decreases
the theoretical cross sections for protons, what causes that
theoretical spectra are below the experimental data for all beam
energies and for all scattering angles.  The height of the
evaporation peak is slightly overestimated in both types of the
calculations.

 Further inspection of Fig. \ref{fig:au234icg} leads to the conclusion
that there are two obvious trends in the difference of the
theoretical spectra evaluated with the coalescence mechanism and the
experimental data: (i) The higher the beam energy, the larger the
underestimation of the high energy part of the data by theory, and
(ii) the smaller the scattering angle in respect to the proton beam
direction, the larger underestimation of the data. This effects
might be explained by the assumption, that an additional process
exists, which manifests itself mainly at small scattering angles and
gives increasing contribution to the emission of protons for larger
beam energies. Such a contribution can correspond to the presence of
the "fireball" emission, which due to fast motion in forward
direction should modify the cross sections mainly at forward
scattering angles. However, in the microscopic calculations
performed according to intranuclear cascade model there is no
explicit room for such a process. Therefore, inclusion of "fireball"
emission should be automatically accompanied by decreasing the
contribution from direct processes simulated by intranuclear cascade
and coalescence of escaping nucleons. According to the reasoning
given to above, the spectra of protons evaluated from intranuclear
cascade with inclusion of coalescence
and with contribution of evaporation of particles 
were multiplied by a factor, common for all scattering angles,
treated as a free parameter of the fit and then added to the
contribution from the "fireball" emission calculated according to
the formula of single moving source emitting isotropically the LCP's
\cite{WES78A}.
The parameters
of the single moving source - the "fireball", i.e. its temperature
parameter - $T$, velocity of the source - $\beta$, total production
cross section associated with this mechanism - $\sigma$ was treated
also as free parameters. Height of the Coulomb barrier between the
"fireball" and emitted ejectile was arbitrarily fixed at 2 \% of the
estimated Coulomb barrier for emission from the target nucleus.
Values of the parameters of "fireball" are given in the Table
\ref{tab:fireball}. 

\begin{table}
 \caption{\label{tab:fireball} Parameters of the "fireball": $\beta$, T,
 and $\sigma$ correspond to "fireball" velocity, its apparent temperature,
 and total (integrated over angle and energy of detected particles)
 production cross section, respectively, B/d determines the ratio of the
 threshold energy for emission of the particles (height of the
 Coulomb barrier) to diffuseness of the transmission function through the barrier.
 Parameter F is the scaling factor of coalescence and evaporation
 contribution extracted from fit to the proton spectra. The numbers in parentheses
 show fixed values of the parameters. Note, that for $\alpha$ particles additional
 moving source should be added with parameters given in the Table \ref{tab:alpha}}
\begin{center}
 \begin{tabular}{llcccccr}
\hline \hline
E$_{p}$  &  Ejectile & $\beta$ &  T    &  $\sigma$ &  B/d  &   F    &  $\chi^{2} $ \\
GeV      &           &         &  MeV  &  mb       &       &        &  \\
\hline
 1.2     & \emph{p}  & 0.136  & 36.7  & 1400     & 11.4  & 0.63  & 27.2 \\
         & \emph{d}  & 0.160  & 39.1  &  190     & 12.1  & [0.63]&  9.5   \\
         & \emph{t}  & 0.073  & 21.5  &   87     &  4.5  & [0.63]&  2.9 \\
         & $^{3}$He  & [0.073]& [21.5]&   0.44   &   18  & [0.63]&  4.5\\
         & $^{4}$He  & 0.070  & 19.0  &   49     &   6.2 & [0.63]&  13.5 \\
\hline
 1.9     & \emph{p}  & 0.160  & 40.7  & 1950     & 11.9  & 0.69   & 46.5 \\
         & \emph{d}  & 0.155  & 41.1  &  330     & 19.0  & [0.69] & 15.3 \\
         & \emph{t}  & 0.066  & 23.8  &  170     &  3.1  & [0.69] & 4.4 \\
         & $^{3}$He  & 0.045  & 15.0  & 15.6     &  5.2  & [0.69] & 3.3 \\
         & $^{4}$He  & 0.061  & 20.9  & 110      &  4.7  & [0.69] & 15.1 \\
\hline
 2.5     & \emph{p}  & 0.156  & 41.7  & 2720     & 12.0  & 0.73   & 39.0 \\
         & \emph{d}  & 0.130  & 42.3  &  530     &  8.6  & [0.73] & 10.5 \\
         & \emph{t}  & 0.050  & 23.3  &  300     &  5.7  & [0.73] &  3.2 \\
         & $^{3}$He  & 0.037  & 20.5  &   54     &  5.8  & [0.73] &  2.7 \\
         & $^{4}$He  & 0.051  & 20.7  &  210     &  3.7  & [0.73] & 11.5 \\
\hline \hline
\end{tabular}
\end{center}
 \end{table}
%

\begin{figure}
\begin{center}
\includegraphics[angle=0,width=0.5\textwidth]{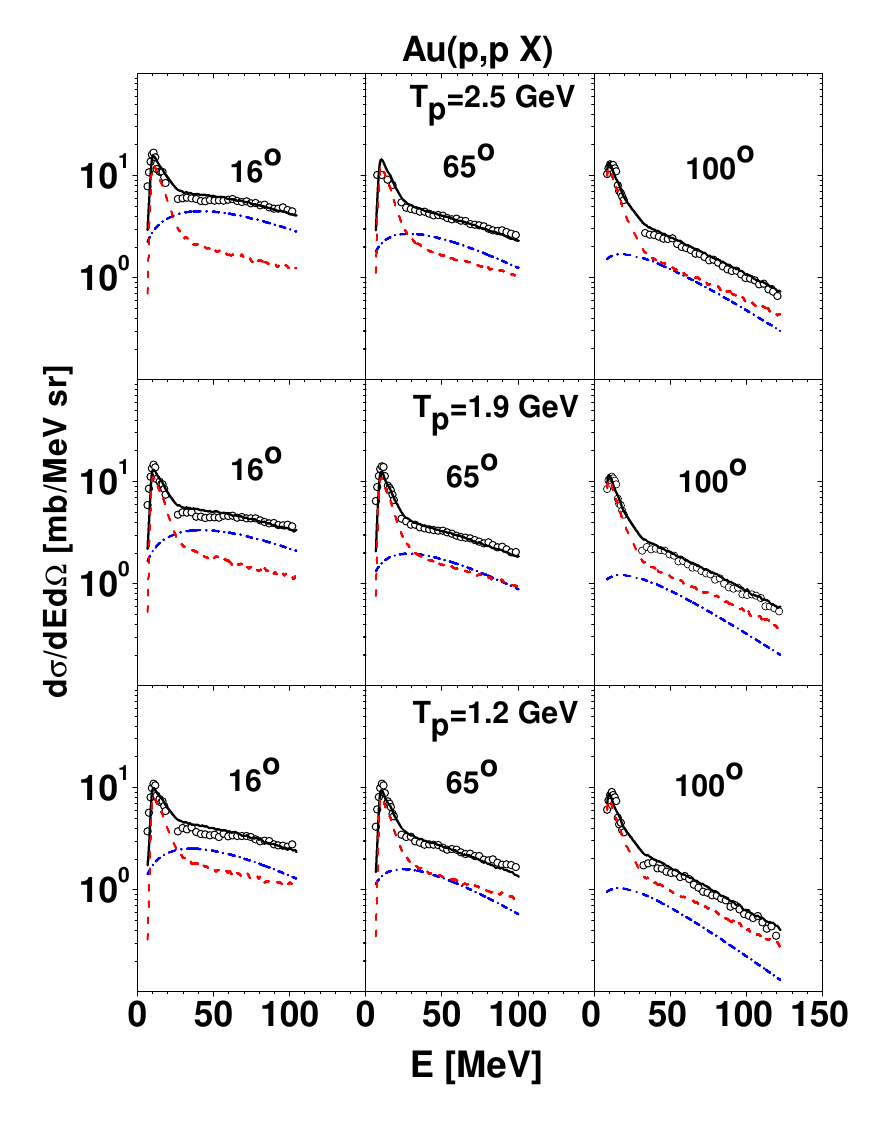}
\caption{\label{fig:pau234icg1s} (Color online) Open circles
represent experimental energy spectra of protons measured at
selected angles: 16$^{\circ}$, 65$^{\circ}$, and 100$^{\circ}$
(left, central, and right column of the figure, respectively)  for
three proton beam energies: 1.2, 1.9 GeV -- present experiment , and
2.5 GeV -- Ref. \cite{BUB07A} (lower, central, and upper row of the
figure, respectively). The dot-dashed lines present the contribution
of proton emission from the "fireball" whereas the dashed lines show
calculations made with INCL4.3 plus GEM2 programs. The INCL4.3 plus
GEM2 contributions are scaled by the factors 0.63, 0.69, and 0.73
for beam energies 1.2, 1.9, and 2.5 GeV, respectively. The solid
lines show sum of all these contributions.
}
\end{center}
\end{figure}

The fit was performed for 7 scattering angles (16$^{\circ}$,
20$^{\circ}$, 35$^{\circ}$, 50$^{\circ}$, 65$^{\circ}$,
80$^{\circ}$, and 100$^{\circ}$). Results of the fit are presented
in Fig. \ref{fig:pau234icg1s} for 3 angles, the smallest, the
intermediate and the largest, where the dashed lines show
contribution of intranuclear cascade with surface coalescence and
evaporation, the dash-dotted lines present contribution from
"fireball" emission, and the solid line depicts sum of both
contributions. As can be seen the excellent agreement could be
obtained for all scattering angles and beam energies. It is worth to
emphasize, that the "fireball" contribution to the spectra increases
significantly both, with the decrease of the scattering angle and
with the increasing of the beam energy.

Success of description of proton spectra by microscopic model of
intranuclear cascade with coalescence of nucleons and evaporation of
protons from equilibrated target residuum combined with
phenomenological contribution from the "fireball" emission shows
that the same method of data description might be applicable for
other LCP's.

It is natural to scale the model coalescence contribution to spectra
of complex LCP's by the same factor "F" which was used for the
proton spectra because  the coalescence emission of complex
particles is determined by the yield of nucleons leaving the nucleus
after intranuclear cascade of collisions.

\begin{figure}
\begin{center}
\includegraphics[angle=0,width=0.5\textwidth]{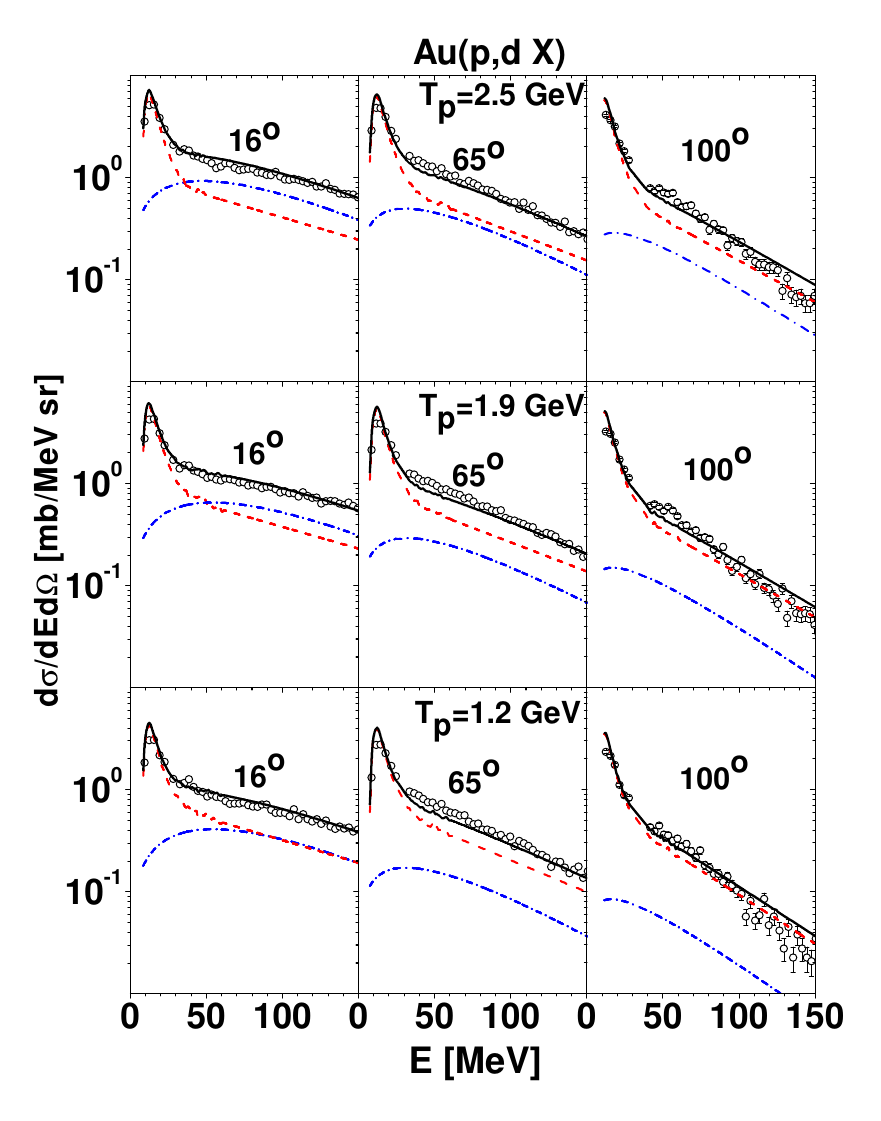}
\caption{\label{fig:dau234icg1s} (Color online) Same as Fig.
\ref{fig:pau234icg1s}, but for deuterons. 
}
\end{center}
\end{figure}

\begin{figure}
\begin{center}
\includegraphics[angle=0,width=0.5\textwidth]{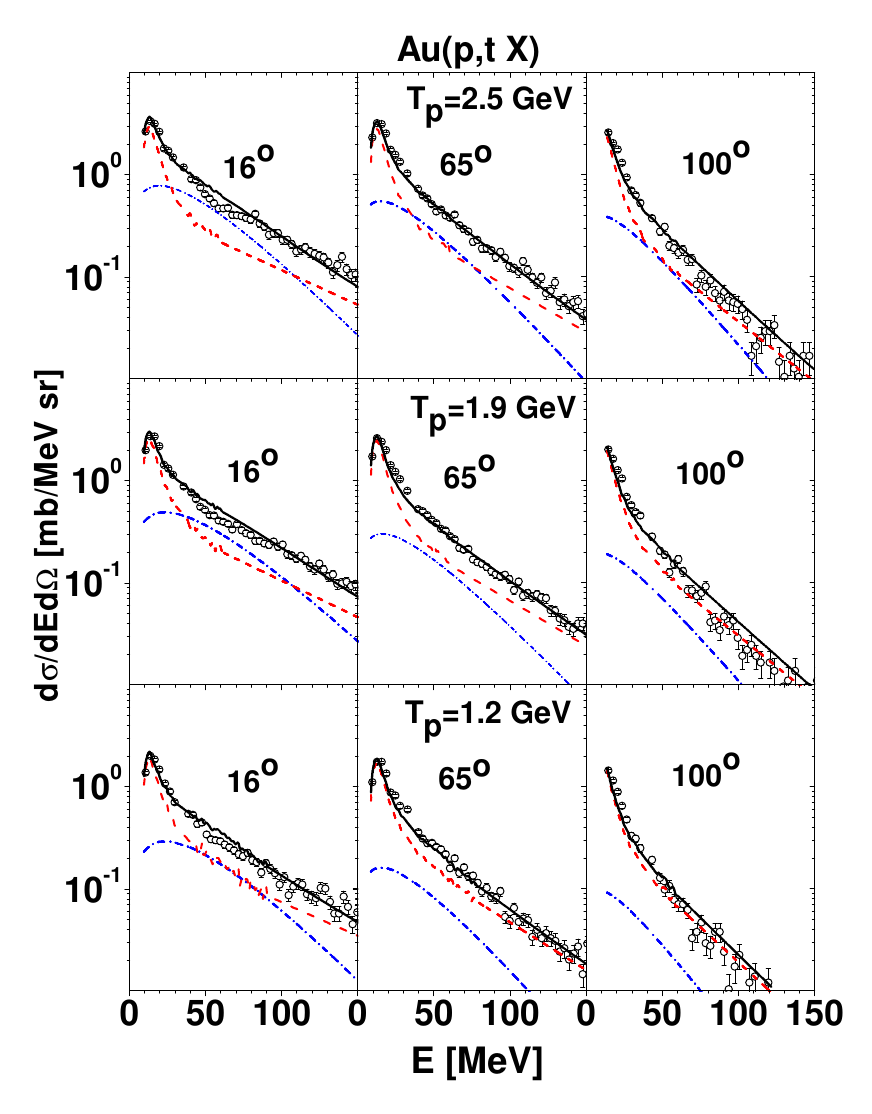}
\caption{\label{fig:tau234icg1s} (Color online) 
Same as Fig. \ref{fig:pau234icg1s}, but for tritons. }
\end{center}
\end{figure}

\begin{figure}
\begin{center}
\includegraphics[angle=0,width=0.5\textwidth]{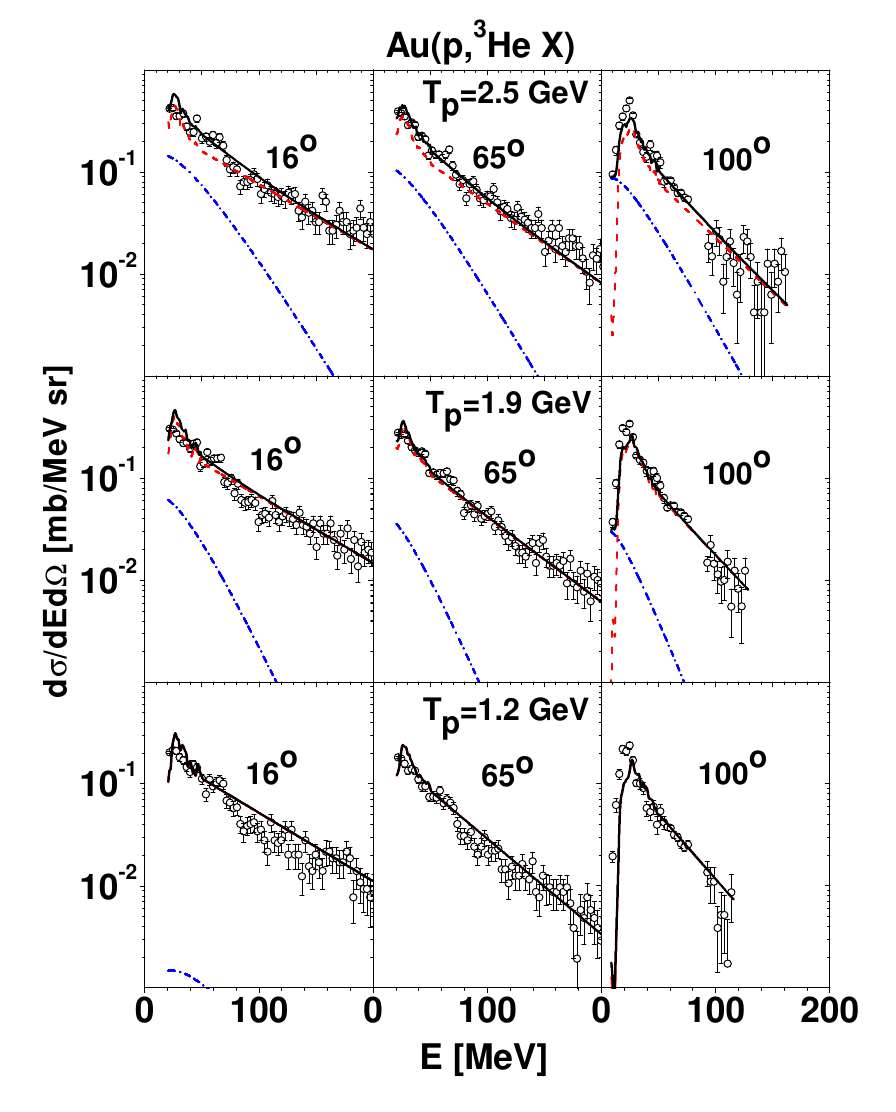}
\caption{\label{fig:hau234icg1s} (Color online) 
Same as Fig. \ref{fig:pau234icg1s}, but for $^{3}$He.}
\end{center}
\end{figure}

\begin{figure}
\begin{center}
\includegraphics[angle=0,width=0.5\textwidth]{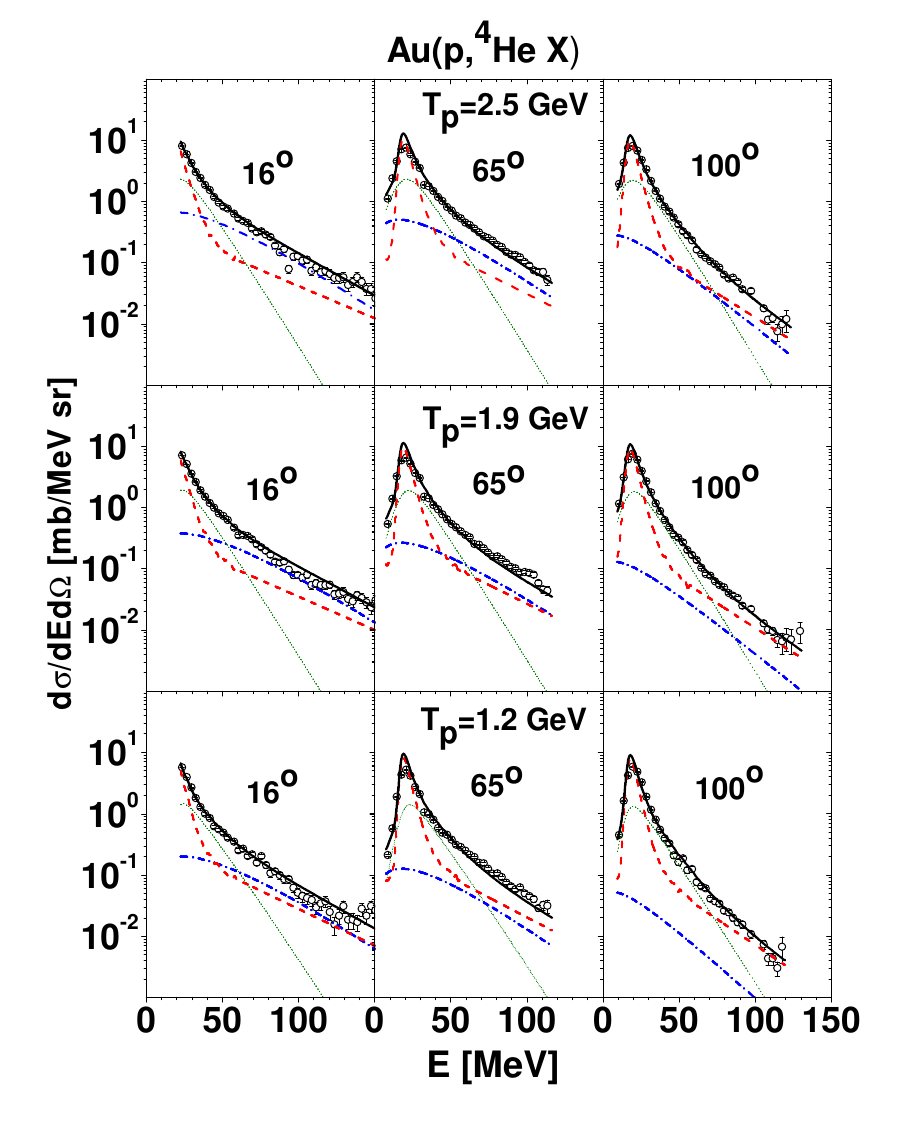}
\caption{\label{fig:aau234icg2s} (Color online) 
Same as Fig. \ref{fig:pau234icg1s}, but for $\alpha$-particles. The
thin dotted line depicts contribution from fast moving source of the
mass intermediate between the "fireball" and the heavy target
residuum. }
\end{center}
\end{figure}

  The fits of parameters
characterizing the "fireball" to the experimental spectra of
deuterons, tritons, $^{3}$He, and $^{4}$He were therefore performed
with the same scaling factors of coalescence and evaporation
emission as those for the proton spectra: 0.63, 0.69, and 0.73 for
beam energy 1.2, 1.9, and 2.5 GeV, respectively. Very good
description of the experimental data was achieved for all particles
with exception of $\alpha$-particles for which it was necessary to
add a contribution of another moving source - with parameters very
close to those used for IMF's. 
This additional contribution led to perfect description of the
$\alpha$ - particle spectra. Quality of the data reproduction is
illustrated by Figs. \ref{fig:dau234icg1s}, \ref{fig:tau234icg1s},
\ref{fig:hau234icg1s}, and \ref{fig:aau234icg2s} for deuterons,
tritons, $^{3}$He, and $^{4}$He, respectively. The parameters of the
"fireball" source are listed in the Table \ref{tab:fireball} and
parameters of additional source used for $\alpha$-particles are
depicted in the Table \ref{tab:alpha}.

As can be seen from the figures, the spectra of deuterons and
tritons could not be described, even qualitatively, by coalescence
and evaporation of particles alone.  The reason of this fact is
difference between angular variation of the experimental spectra and
those evaluated from the microscopic model.  For example,
multiplication of coalescence spectra by factor which allows to well
reproduce spectrum at 100$^{\circ}$ still leads to underestimation
of the cross sections at smaller angles. On the contrary, adding the
contribution of emission of deuterons and tritons from the
"fireball" improves the description significantly because this
contribution has exactly such an angular and energy dependence which
added to microscopic model spectra assures reproduction of the
experimental data.

A different situation is present for $^{3}$He channel, where the
"fireball" contribution seems to be almost negligible, especially at
lower beam energies.  It means, that the coalescence together with
small evaporation contribution exhaust almost fully the experimental
yield of particles leaving no room for the "fireball" emission.
%
%
%
It should be, however, emphasized that this very good data
reproduction by the coalescence and evaporation mechanisms was
obtained after scaling of the theoretical cross sections from
INCL4.3+GEM2 by the same factors as those used for the theoretical
cross sections for proton emission, thus the presence of "fireball"
emission influences also indirectly the description of $^{3}$He
emission.

Still another reaction mechanism seems to be responsible for the
$\alpha$-particle production.  The shape as well as magnitude of the
experimental spectra for $^{3}$He and $^{4}$He is quite different,
showing that evaporation of $\alpha$-particles from excited target
residuum after intranuclear cascade of nucleon-nucleon collisions is
much more abundant than corresponding evaporation of $^{3}$He
particles. However, the peak present in the experimental spectra of
$^{4}$He is much broader than that predicted by evaporation from
heavy target residuum.  Since neither coalescence mechanism nor
"fireball" emission can produce such a peak in the spectrum, thus,
another contribution is necessary to reproduce  the shape of the
peak in the experimental
spectra.  
The naturally appearing so\-lu\-tion is to take into consideration
the contribution from the moving source of the mass larger than the
"fireball" but smaller than heavy target residuum. Such a source,
moving faster than target residuum but slower than the "fireball",
was observed in the analysis of spectra for all IMF's, thus it is
not astonishing that also $\alpha$-particle spectra are modified by
its contribution.

%


\section{\label{sec:discussion}Discussion}

\begin{figure}
\begin{center}
\includegraphics[angle=0,width=0.5\textwidth]{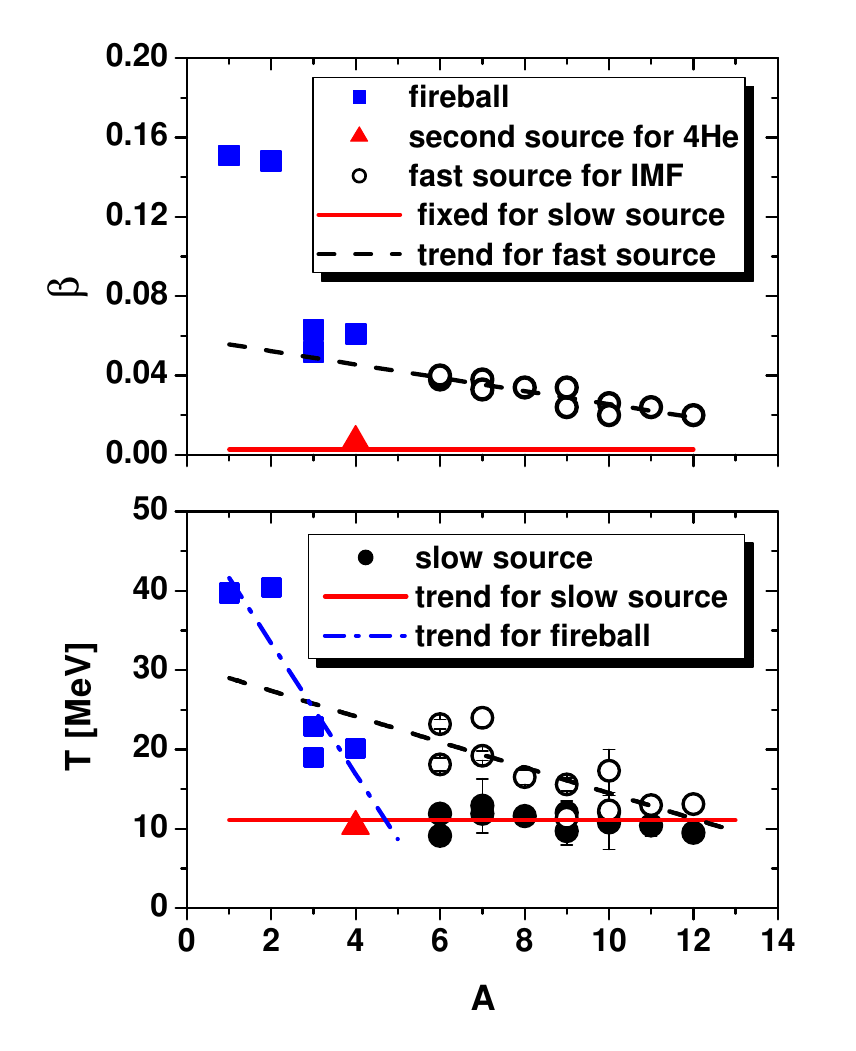}
\caption{\label{fig:bt123s} (Color online) In the \emph{lower par}t
of the figure the apparent temperature of the moving sources,
averaged over beam energies is drawn as a function of the ejectile
mass. Open circles and full dots represent values of parameters
obtained from analysis of IMF's data for fast and slow source,
respectively. Full squares depict temperature of the "fireball"
fitted to spectra of LCP's together with the contribution of
microscopic model of intranuclear cascade, coalescence of nucleons
and statistical evaporation. Full triangle shows the temperature of
the second source fitted to spectra of $\alpha$-particles. The solid
and dashed lines were fitted to the points representing IMF's and
extrapolated to smaller masses. Dash dotted line was fitted to LCP's
temperatures of the "fireball". In the \emph{upper part} of the
figure the dependence of the beam energy averaged velocity of the
sources is drawn versus mass of ejectiles. The symbols have the same
meaning as for the lower part of the figure with one exception: The
full dots are not shown because the velocity of slower source was
fixed during analysis (at velocity of heavy residuum of target
nucleus after intranuclear cascade) and it is represented by solid
line in the figure. The dashed line was fitted to open circles
representing velocities of fast source for IMF's. The line was
extrapolated to lower mass region.}
\end{center}
\end{figure}

The temperature of the "fireball" fitted to describe LCP's data
varies only slightly with the beam energy.  Its values listed in the
Table \ref{tab:fireball} do not change more than $\sim$ 10$\%$ for
each ejectile in the beam energy range from 1.2 to 2.5 GeV. This is
also true for the temperature of the additional source necessary to
be included for good description of $\alpha$-particle data and for
temperatures of both phenomenological sources applied for
parametrization of IMF's data.  This fact allows to study dependence
of the beam energy averaged temperature on the ejectile mass instead
temperature dependencies for individual beam energies.
Beam energy averaged temperatures of all moving sources are depicted
in the lower part of Fig. \ref{fig:bt123s} as function of the
ejectile mass A. It is seen that temperatures of two sources
emitting IMF's are grouped into two sets: the full dots -
representing slow sources - lie along solid horizontal line $T=11.1$
MeV whereas the open circles - representing fast sources - are
spread around the dashed line $T=30.6-1.61 A$ MeV. The same
procedure applied to apparent temperatures of the "fireball"
emitting LCP's shows that the mass dependence of this temperature
may be described by linear function: $T=49.9-8.24 A$ MeV (dash
dotted line in Fig. \ref{fig:bt123s}).

If the ejectile mass A dependence of the apparent temperature T of
the source is caused only by recoil of the source during emission of
registered ejectiles then it is possible to estimate mass of the
source $A_{S}$ and its true temperature $\tau$ from parameters of
the linear dependence T(A).  For the fast source emitting IMF's the
source temperature is equal to $\tau$=30.6 MeV and mass of it is
equal to A$_{S}$=30.6/1.61 $\equiv$ 19 nucleons. The temperature of
the slow source is independent of the IMF's mass what means that the
recoil effect is negligible, i.e. the source is very heavy and its
temperature is equal to apparent temperature found in the fit
($\tau$=11.1 MeV). The temperature of "fireball" extracted from the
parameters of the fitted straight line is equal to $\tau$=49.9 MeV
and the "fireball" is built of A$_{S}$=49.9/8.24 $\equiv$ 6
nucleons.

These conclusions seem to be compatible with results of pure
phenomenological analysis of two moving sources performed in our
previous investigation of LCP's and IMF's for Au+p collisions at
proton beam energy 2.5 GeV \cite{BUB07A}. In this study the
temperature of the slow source for IMF's was $\sim$ 12 MeV, the
temperature of the fast source for IMF's was $\sim$ 33 MeV, and the
temperature of "fireball" was estimated to be $\sim$ 62 MeV.  Mass
of the slow source must be very large - close to the mass of the
target - because apparent temperature of this source did not vary
significantly with the product mass, i.e. recoil could be neglected.
The mass of the fast source was equal to mass of $\sim$~20 nucleons
and mass of the "fireball" was close to the mass of $\sim$~8
nucleons.

The largest deviation between previous results and those found in
the present work concern properties of the "fireball". This is not
surprising because the "fireball" of the present work is responsible
only for a part of the effect which was attributed to the "fireball"
in the previous study. However, inspection of Fig. \ref{fig:bt123s}
shows also another effect: The straight dashed line representing
apparent temperature of the fast source with the mass of about 19
nucleons - found from analysis of IMF's data - crosses  the dash
dotted line representing apparent temperature of the "fireball" at
mass of ejectile A $\sim$ 3.  It means that the temperature
parameter of the "fireball" and that of the intermediate mass source
are the same for tritons, $^{3}$He, and $^{4}$He. Moreover, the
velocity of the "fireball" emitting tritons, $^{3}$He, and $^{4}$He
is very close to velocity of the fast source emitting IMF's as it is
shown in the upper part of Fig. \ref{fig:bt123s} where the beam
energy averaged values of the velocity parameter are collected for
IMF's (open circles for the fast source and solid, horizontal line -
fixed at velocity of heavy residuum from intranuclear cascade - for
the slow source) and for LCP's (full squares for the "fireball" and
the full triangle for additional source necessary for description of
the $\alpha$-particles).
Thus, it is not clear whether it is allowed to extract mass of the
"fireball" from mass dependence of the apparent temperature of the
source fitted to proton, deuteron, triton, and $^{3,4}$He data or it
is necessary to assume that the source for particles with mass 3 and
4 is identical with the intermediated mass source (A$_S \sim$ 19)
found for IMF's. If this is the case, then the \emph{genuine}
"fireball" contributes mainly to emission of protons and deuterons,
thus it is reasonable to conjecture that the mass of the "fireball"
should be very light (3-4 nucleons).


\begin{table}
\caption{\label{tab:alpha} Parameters of the intermediate mass
source needed to describe well the $\alpha$ - particle spectra by
combination of microscopic model coalescence and evaporation
contributions, the "fireball" and intermediate mass source
contributions. Parameters $\beta$, T, and $\sigma$ have the same
meaning as that given in Table \ref{tab:fireball} for the
"fireball". The k parameter is the height of the Coulomb barrier in
units of simple barrier height estimation by Coulomb potential of
two uniformly touching spheres with the charge of the target nucleus
and the charge of the emitted particle with radii parameterized as
R=1.44 A$^{1/3}$.  }
\begin{center}
\begin{tabular}{lllll}
\hline \hline

E$_{p}$   & k               & $\beta$             &  T            &  $\sigma$    \\
GeV       &                 &                     &  MeV          &  mb           \\
\hline
1.2       & [0.8]           & 0.0094              & 10.6          &  385  \\
1.9       & 0.83$\pm$0.03   & 0.0062$\pm$0.0010   & 10.2$\pm$0.3  &  577$\pm$23 \\
2.5       & 0.80$\pm$0.04   & 0.0047$\pm$0.0011   & 10.2$\pm$0.4  &  764$\pm$38 \\
\hline \hline
\end{tabular}
\end{center}
\end{table}

\begin{table}
\caption{\label{tab:coalescence} Cross sections (in millibarns) for
production of LCP's by intranuclear cascade and coalescence
mechanism (left part of the table), and by the evaporation (right
part of the table) - evaluated with INCL4.3 + GEM2 computer programs
and scaled by appropriate factors: 0.63, 0.69, and 0.73 for beam
energies 1.2, 1.9, and 2.5 GeV, respectively. }
\begin{center}
\begin{tabular}{lrrrrrrrrrr}
\hline \hline

E$_{p}$   &  \multicolumn{5}{c}{coalescence}  & \multicolumn{5}{c}{evaporation}    \\
GeV       &    p  &   d    &    t   & $^{3}$He & $^{4}$He &   p  & d   & t  & $^{3}$He & $^{4}$He   \\
\hline
1.2       & 2213  &  613   &  198   &   75  &   62        & 633  & 272 & 134 &  10.2   & 718 \\
1.9       & 2740  &  771   &  254   &  101  &   80        & 932  & 432 & 212 &  20.4   & 914 \\
2.5       & 3084  &  859   &  285   &  116  &   90        & 1094 & 526 & 257 &  27.1   & 1019 \\
\hline \hline
\end{tabular}
\end{center}
\end{table}

%

It is worth to point out that values of temperature and velocity of
the additional source introduced to describe the $\alpha$-particle
emission (triangles in Fig. \ref{fig:bt123s}) are very similar to
values characterizing the slow, heavy source emitting the IMF's
(solid line in Fig. \ref{fig:bt123s}.

All these findings agree well with conclusions derived from pure
phenomenological analysis of the p+Au data measured at 2.5 GeV
proton beam energy \cite{BUB07A}, which consist in the statement,
that nonequilibrium contribution to production of LCP's and IMF's
indicates presence of the mechanism similar to fast break up of the
target nucleus 
in which three moving sources of ejectiles are created.  The new
result of the present work is an observation that the nonequilibrium
emission of LCP's is mediated by two competing mechanisms: surface
coalescence of outgoing nucleons and the contribution from three
moving sources appearing as result of the break up.


It is interesting to compare cross sections for inclusive LCP's
production originating from these two nonequilibrium mechanisms.
Values of cross sections for nonequilibrium processes are listed in
the Tables \ref{tab:fireball}, \ref{tab:alpha} and
\ref{tab:coalescence}, for emission from the "fireball", for
emission from additional, slower source, and from coalescence,
respectively. Proton beam energy dependence of these cross sections
as well as dependence of relative contribution of the "fireball"
emission are presented in Fig. \ref{fig:cfrode}. Several important
conclusions can be derived from inspection of the figure:
\begin{itemize}
  \item  Cross sections for all emitted LCP's increase with
            energy in approximately exponential way, however, this increase is
            faster for the "fireball" emission (central part of the figure) than
            for the coalescence mechanism (lower part of the figure).
  \item  Magnitude of the coalescence cross sections decreases
         strongly with the mass of ejectile - cf. values of the cross sections
         in the lower part of Fig. \ref{fig:cfrode}: squares - for protons,
         dots - for deuterons, triangles - for tritons, stars connected by
         straight lines - for $^{3}$He, and diamonds - for $\alpha$-particles.
         This behavior may be explained by
         decreasing probability of capture of more and more nucleons by the nucleon
         escaping from the nucleus.
\item    The cross sections for two
         isobars - triton and $^{3}$He are quite different.
         The cross section for triton production is approximately two times
         larger than that for $^{3}$He.
         Such a big difference may be related to ratio N/Z=1.49 of Au nuclei
         and may be additionally enhanced by the fact that
         coalescence of two neutral particles - neutrons and one charged particle - proton
         is not influenced by repulsive Coulomb force, whereas coalescence of two protons
         and one neutron is certainly hindered to some extent by Coulomb interaction.
         This effect is also visible for "fireball" emission of tritons and $^{3}$He,
         moreover, the ratio of triton cross section to $^{3}$He
         cross sections is larger than for coalescence and varies (decreases)
         strongly with beam energy - cf. central part of Fig. \ref{fig:cfrode}.

\begin{figure}
\begin{center}
\includegraphics[angle=0,width=0.5\textwidth]{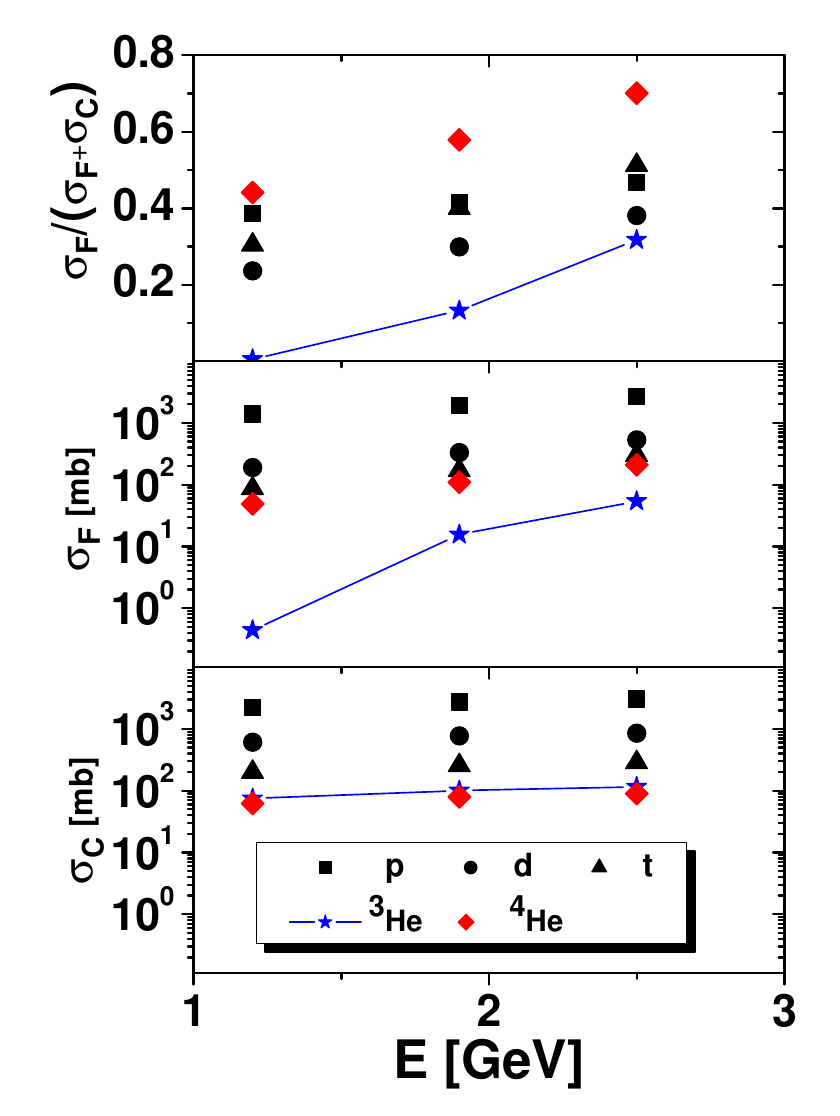}
\caption{\label{fig:cfrode} (Color online) In the upper part of the
figure the beam energy dependence of the relative contribution of
"fireball" emission to the whole nonequilibrium production cross
sections is shown. In the central part of the figure the energy
dependence of the cross section due to the "fireball" mechanism and
in the lower part of the figure the energy dependence of the
production cross sections due to coalescence are depicted. Full
squares, dots, and triangles represent proton, deuteron and triton
cross sections, respectively. The stars connected by solid line show
cross sections for $^{3}$He production and the diamonds correspond
to $\alpha$-particle cross sections. }
\end{center}
\end{figure}

%
  \item Relative contribution of "fireball" mechanism increases almost exponentially with
        the beam energy - cf. upper part of Fig. \ref{fig:cfrode}.  Slope of the energy dependence
        is smallest for protons, has intermediate value for deuterons and tritons, and
        is largest for $^{3}$He and $\alpha$-particles.  It should
        be, however, pointed out that in spite of such a fast increase of
        "fireball" emission for both helium isotopes, the contribution
        of this mechanism is less important for these particles than
        for the hydrogen isotopes.  For $^{3}$He this is caused by the
        fact that relative contribution of "fireball" emission is
        small - smaller than 20\% of the sum of  both considered mechanisms,
        and furthermore, it manifests
        itself only at forward angles and small energies (cf. Fig.
        \ref{fig:hau234icg1s}). Thus, $^{3}$He spectra can be quite well
        reproduced by scaled coalescence mechanism contribution alone.
        For the $\alpha$-particles, the
        "fireball" contribution is comparable to that of the
        coalescence mechanism, however, as it was discussed above,
        another nonequilibrium process gives large contribution to
        the experimental spectra: emission from the source of mass
        intermediate between the "fireball" and the heavy target
        residuum created in the fast stage of the reaction.
\end{itemize}

\begin{figure}
\begin{center}
\includegraphics[angle=0,width=0.5\textwidth]{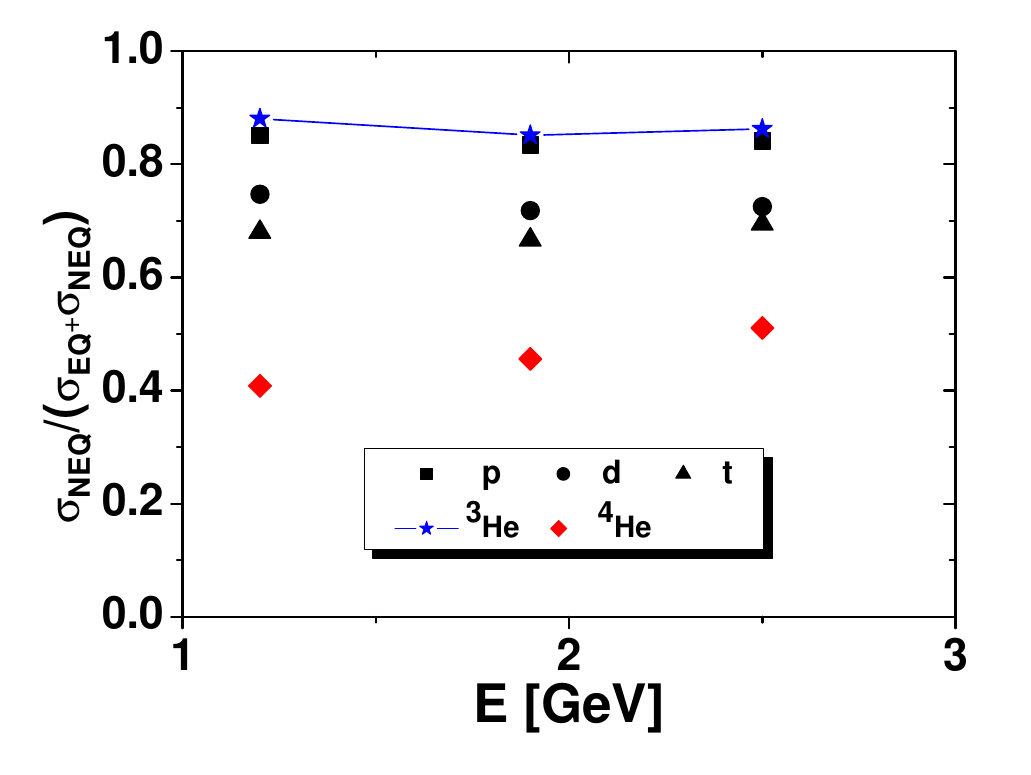}
\caption{\label{fig:rneqtosum} (Color online) The ratio of the sum
of cross sections for nonequilibrium processes (the coalescence and
"fireball" emission cross sections for protons, deuterons, tritons,
and $^{3}$He particles, whereas for $\alpha$-particles still cross
section for emission from intermediate mass source is added) to sum
of cross sections for all these processes and cross section for
emission from the equilibrated target residuum after intranuclear
cascade is shown as function of beam energy.  The  cross sections
evaluated by programs INCL4.3+GEM2 were multiplied by factors 0.63,
0.69, and 0.73 for beam energies 1.2, 1.9, and 2.5 GeV,
respectively. The same symbols are used as in Fig. \ref{fig:cfrode},
i.e., full squares present results for protons, full dots - results
for deuterons, triangles - correspond to tritons, stars connected by
solid line depict the $^{3}$He data and full diamonds represent
$\alpha$-particles.}
\end{center}
\end{figure}

The present investigations allowed to find the beam energy variation
of the contribution of nonequilibrium processes to the studied
emission of LCP's. In Fig. \ref{fig:rneqtosum} the ratio is shown of
the sum of all nonequilibrium processes, i.e., the "fireball"
emission and the coalescence for p, d, t, and $^{3}$He ejectiles
with additional contribution of the intermediate mass source for
$^{4}$He particles, to sum of all these processes and compound
nucleus cross section evaluated by INCL4.3+GEM2 programs and scaled
by factors found from the fit to the proton spectra. As can be seen
from the figure the contribution of nonequilibrium processes is very
large for all energies.  It has the largest values (over 80 \%) for
$^{3}$He and for protons.  For deuterons and tritons this
contribution is about 70~\%, whereas for $\alpha$-particles it is
the smallest, however, still quite large (40 - 50~\% - depending on
the beam energy). The important conclusion is that the energy
dependence of the relative contribution of nonequilibrium processes
is very weak with the exception of the $\alpha$-particles, where
this relative contribution increases by 20\% from the lowest beam
energy to the highest one. Such - almost constant - value of the
contribution of nonequilibrium processes seems to be rather
unexpected in view of strong increase of the total production cross
sections in the studied, broad range of the beam energy. However, it
may be explained by the fact that cross sections of both,
equilibrium and nonequilibrium processes, increase with energy in similar manner.\\

\section{\label{sec:summary}Summary and conclusions}

Double differential cross sections $d^{2}\sigma/d\Omega dE$ 
were measured for $p,d,t,^{3,4,6}$He, $^{6,7,8,9}$Li, $^{7,9,10}$Be,
and $^{10,11,12}$B  produced in collisions of 1.2 and 1.9 GeV
protons with  Au target. It was found that the spectra measured at
16, 20, 35, 50, 65, 80, and 100 degrees in the present experiment as
well as such data obtained at 2.5 GeV beam energy \cite{BUB07A} are
very similar indicating large contribution of nonequilibrium
processes. The data for IMF's were analyzed in the frame of
phenomenological model of two moving sources emitting isotropically
the ejectiles. The slow source simulated evaporation of particles
from the equilibrated remnant of the intranuclear cascade of
nucleon-nucleon collisions whereas the fast source was responsible
for description of nonequilibrium processes.  Very good reproduction
of all cross sections was achieved with the parameters which vary
smoothly with the beam energy and mass of ejectiles. It was found
that the cross sections corresponding to both sources increase with
the energy and the relative contribution of the nonequilibrium
processes varies from (27 $\pm$ 3)$\%$ at 1.2 GeV beam energy to (44
$\pm$ 5)$\%$ at 2.5 GeV beam energy.

The LCP's data were analyzed by means of the microscopic model which
takes into consideration the intranuclear cascade of nucleon-nucleon
collisions, coalescence of the nucleons escaping from the nucleus
after the cascade, and the evaporation of particles from the
equilibrated, excited residuum of the target nucleus. The
calculations were performed using the computer program INCL4.3 of
Boudard et al. \cite{BOU04A} for intranuclear cascade  and
coalescence processes, and by GEM2 computer program of S.Furihata
\cite{FUR00A},\cite{FUR02A} for evaporation.  It should be
emphasized that free parameters of both models have been not fitted
to the data but original values of these parameters, recommended by
the authors, have been used.
Model cross sections were significantly smaller than the
experimental data for emission of protons with energies larger than
$\sim$ 30 MeV whereas the evaporation contribution, which dominates
the smaller energy range of spectra, overestimates the data.  The
discrepancy increases with increasing beam energy and with
decreasing the emission angle.

It was assumed that an additional contribution to the microscopic
model cross sections has to be added to account for the observed
discrepancies in the description of proton data. The isotropic
emission of particles from the "fireball" moving forward, i.e.,  in
the direction parallel to the beam, leads to desirable energy and
angular distributions. Thus, this process has been taken into
consideration for improving the proton data description. Parameters
of the "fireball" were treated as free parameters. The magnitude of
the contribution from microscopic model was allowed to be scaled
down because of two reasons: (i) in the intranuclear cascade model
it is assumed that \textbf{\emph{each}} proton bombarding the
nucleus initiates the cascade of nucleon-nucleon collisions, thus,
"fireball" process, which consists in creation of \textbf{\emph{a
correlated group}} of the nucleons emitted in the forward direction,
is completely neglected, (ii) the magnitude of the coalescence
process may be modified by variation of the conditions which
determine whether nucleons form the cluster or  move independently.

Excellent description of the proton spectra was achieved for all
emission angles and for all beam energies with the parameters of the
"fireball" 
varying smoothly with the
beam energy. Furthermore, the factor which was used to scale down
the contribution from intranuclear cascade modified by coalescence
and the contribution of the evaporation was almost energy
independent: 0.63, 0.69, and 0.73 for beam energy 1.2, 1.9, and 2.5
GeV respectively.

The spectra of other LCP's were analyzed in the same manner, i.e.,
the microscopic model contribution (coalescence and evaporation
cross sections) was multiplied by the same factor which was used for
the proton channel and parameters of the "fireball" were fitted
independently for each ejectile and each beam energy. Excellent
description of all data has been obtained with smoothly varying
parameters of the "fireball". The data for $^{4}$He channel still
need inclusion of the contribution from another slow  moving
source. 

The  contribution of the "fireball" mechanism to the nonequilibrium
processes is quite significant for all light charged particles (20\%
- 60\% - depending on the particles and beam energy). Magnitude of
this contribution
increases almost exponentially with the beam
energy.

Rather astonishing result of the present investigation, that the
relative contribution of all nonequilibrium processes to the total
cross sections (40\% - 80\%, depending on the particles) remains
almost energy independent for all light charged particles is caused
by presence of similar energy dependence for both, equilibrium and
nonequilibrium processes in the studied energy range.  Such a weak
energy dependence of the relative contribution of nonequilibrium
processes was also found for production of intermediate mass
fragments, as it was stated above.

Comparison of parameters of moving sources used in description of
IMF's and LCP's data at three proton energies; 1.2, 1.9 and 2.5 GeV
confirms our hypothesis postulated in Ref. \cite{BUB07A}, which
claims that the proton impinging on the Au target interacts with
group of nucleons lying on its straight way through the nucleus what
leads to emission of a "fireball" consisted of several nucleons. The
excited remnant nucleus may decay into two prefragments which
manifest themselves as moving sources of LCP's and IMF's, whereas
the "fireball" emits only LCP's.  It was found in the present
analysis that the parameters of the "fireball" fitted to spectra of
tritons, $^{3}$He and for $\alpha$-particles are very similar to
parameters of the light source emitting IMF's.  Therefore, it seems
that the \textbf{\emph{genuine}} "fireball" contributes mainly to
emission of protons and deuterons and, thus, it is consisted of 3 -
4 nucleons. Then the lighter prefragment (of mass of $\sim$ 19
nucleons), appearing as result of decay of excited remnant, is
responsible for emission of tritons, $^{3}$He and
$\alpha$-particles.  The spectra of $\alpha$-particles show also
large contribution originating from the larger prefragment, i.e.
from the emission of slow source responsible for IMF's production.

These findings are in agreement with observations made for hadron
production in high energy (of order of 50 - 200 GeV) proton-nucleus
collisions \cite{BER76A},\cite{HAL77A},\cite{MEN78A} where the
reaction does not proceed on the total nucleus of mass A but the
bombarding proton interacts with \textbf{\emph{the effective
target}} consisted of several nucleons; $\sim 0.7 A^{0.31}$
\cite{HAL77A}. For the Au target such an effective target would have
a mass of 3.6 nucleons, what fits well with estimated mass of the
fireball of the present study and justifies identification of the
"fireball" with the "effective target". Furthermore, the deep
spallation process of production of $^{149}$Tb, studied by Winsberg
\emph{et al.} \cite{WIN76A} in proton - Au collisions at energies 1
- 300 GeV was also explained by Cumming \cite{CUM80A} assuming
manifestation of the effective target with the mass of
(3.1~$\pm$~0.4) nucleons for proton energies larger than $\sim$ 2
GeV.  This mass again agrees with the "fireball" mass found in our
investigations and confirms proposed interpretation of the
"fireball".

It is worth to point out that the observation of the effective
target was also reported in production of heavy fragments in proton
- U collisions at proton energies 11.5 GeV \cite{WIL79A} (A= 140 -
210), \cite{BIS79A} (A=131), and light fragments at 11.5 - 400 GeV
\cite{FOR80B}($^{44}$Sc - $^{48}$Sc) , as well as IMF's (with Z= 3 -
14) in proton - Xe collisions at 1 - 19 GeV energies \cite{POR89A}.

The presence of heavier sources accompanying the "fireball"  and
mechanism of their creation was predicted and discussed in Refs.
\cite{AIC84A}, \cite{CIR81A}, \cite{BOH83A}, and \cite{HUF83A} as
result of "cleveage" of the excited remnant nucleus into two excited
prefragments after "fireball" emission.  In our analysis these
heavier prefragments manifest themselves as two sources emitting
IMF's as well as tritons, $^{3}$He and $\alpha$-particles. Their
contribution to proton and deuteron spectra is not pronounced, thus
it seems, that  the proton and deuteron spectra are dominated by
emission from fireball (large ejectile energies) and by evaporation
from heavy target residuum (small ejectile energies).

In summary, our investigations lead to a consistent picture of
reaction mechanism responsible for nonequilibrium processes, in
which the proton impinging on the target can either initiate cascade
of binary nucleon-nucleon collisions accompanied by surface
coalescence of nucleons into LCP's or interacts coherently with a
group of nucleons leading to emission of three excited groups of
nucleons; the "fireball" and two heavier prefragments with different
masses. All three excited groups of nucleons are sources of
ejectiles. Present investigation shows, that the presence of the
effective target and - in consequence - the fast break up mechanism,
manifests itself at proton beam energies 1.2 - 2.5 GeV, lower than
those from previous studies, quoted to above.

The important conclusion of the present study is the statement, that
for good description of the double differential cross sections for
all LCP's 
it is necessary to assume competition of two mechanisms of the
nonequilibrium processes: coalescence of nucleons escaping from the
nucleus after intranuclear cascade of nucleon-nucleon collisions and
isotropic emission of LCP's from the fast source - "fireball" -
moving forward along the beam direction. A need to introduce
presence of the "fireball" contribution seems to indicate that the
lack of correlation between nucleons, inherent in intranuclear
cascade models, leads to oversimplified microscopic description of
the reaction mechanism. Thus, the realistic microscopic model has to
take this effect into
consideration.\\

\begin{acknowledgments}
We acknowledge gratefully the fruitful discussions on the
coalescence mechanism with A.Boudard, J.Cugnon, and S. Leray as well
as providing us with new version of INCL4.3 computer program. The
technical support of A.Heczko, W. Migda{\l}, and N. Paul in
preparation of experimental apparatus is greatly appreciated. This
work was supported by the European Commission through European
Community-Research Infrastructure Activity under FP6 "Structuring
the European Research Area" programme (CARE-BENE, contract number
RII3-CT-2003-506395 and Hadron Physics, contract number
RII3-CT-2004-506078) as well as the FP6 IP-EUROTRANS
FI6W-CT-2004-516520.
\end{acknowledgments}

\end{document}